\documentclass[%
aps,nofootinbib,notitlepage,superscriptaddress,twocolumn,10pt,prd,
 amsmath,amssymb
]{revtex4-2}

\usepackage[caption=false]{subfig}
\usepackage{braket}
\usepackage{float}
\usepackage{lipsum}
\usepackage{graphicx}
\usepackage{dcolumn}
\usepackage{bm}
\usepackage{natbib}
\usepackage{hyperref}
\hypersetup{
	colorlinks = true,
    linkcolor = Maroon,
    urlcolor  = Maroon,
    citecolor = Maroon
}

\usepackage{microtype}

\usepackage{verbatim}
\usepackage[amssymb]{SIunits}
\usepackage{tabularx}
\usepackage[dvipsnames]{xcolor}

\usepackage{tikz}

\newcommand{\revision}[1]{{\color{black}#1}}

\usepackage{color}
\definecolor{darkgreen}{RGB}{0,120,0}

\newcommand{\vk}{{\bm k}}
\newcommand{\vq}{{\bm q}}
\newcommand{\vp}{{\bm p}}
\newcommand{\av}[1]{\langle{#1}\rangle}

\begin{document}

\title{Squeezing \texorpdfstring{$\bm f_{\rm NL}$}{}  out of the matter bispectrum with consistency relations} 

\author{Samuel~Goldstein}
\email{sjg2215@columbia.edu}
\affiliation{Center for Theoretical Physics, Department of Physics,
Columbia University, New York, NY 10027, USA}
\affiliation{Max-Planck-Institute f\"ur Astrophysik, Karl–Schwarzschild–Stra\ss e 1, 85748 Garching, Germany}

\author{Angelo~Esposito}
\affiliation{School of Natural Sciences, Institute for Advanced Study, 1 Einstein Drive, Princeton, NJ 08540, USA}
\affiliation{Dipartimento di Fisica, Sapienza Universit\`a di Roma, Piazzale Aldo Moro 2, I-00185 Rome, Italy}

\author{Oliver~H.\,E.~Philcox}
\affiliation{Center for Theoretical Physics, Department of Physics,
Columbia University, New York, NY 10027, USA}
\affiliation{Simons Society of Fellows, Simons Foundation, New York, NY 10010, USA}
\affiliation{Department of Astrophysical Sciences, Princeton University, Princeton, NJ 08540, USA}
\affiliation{School of Natural Sciences, Institute for Advanced Study, 1 Einstein Drive, Princeton, NJ 08540, USA}

\author{Lam~Hui}
\affiliation{Center for Theoretical Physics, Department of Physics,
Columbia University, New York, NY 10027, USA}

\author{J.\,Colin~Hill}
\affiliation{Center for Theoretical Physics, Department of Physics,
Columbia University, New York, NY 10027, USA}
\affiliation{Center for Computational Astrophysics, Flatiron Institute, New York, NY 10010, USA}

\author{Roman~Scoccimarro}
\affiliation{Center for Cosmology and Particle Physics, Department of Physics, New York University, New York, NY 10003, USA}

\author{Maximilian~H.\,Abitbol}
\affiliation{University of Oxford, Department of Physics, Denys Wilkinson Building, Keble Road, Oxford OX1 4LS, UK}

\begin{abstract}
    We show how consistency relations can be used to robustly extract the amplitude of local primordial non-Gaussianity ($f_{\rm NL}$) from the squeezed limit of the matter bispectrum, well into the non-linear regime. First, we derive a non-perturbative relation between primordial non-Gaussianity and the leading term in the squeezed bispectrum, revising some results present in the literature. This relation is then used to successfully measure $f_{\rm NL}$ from $N$-body simulations. We discuss the dependence of our results on different scale cuts and redshifts. Specifically, the analysis is strongly dependent on the choice of the smallest soft momentum, $q_{\rm min}$, which is the most sensitive to primordial bispectrum contributions, but is largely independent of the choice of the largest hard momentum, $k_{\rm max}$, due to the non-Gaussian nature of the covariance. We also show how the constraints on $f_{\rm NL}$ improve at higher redshift, due to a reduced off-diagonal covariance. In particular, for a simulation with $f_{\rm NL} = 100$ and a volume of $(2.4 \text{ Gpc}/h)^3$, we measure $f_{\rm NL} = 98 \pm 12$ at redshift $z=0$ and $f_{\rm NL} = 97 \pm 8$ at $z=0.97$. Finally, we compare our results with a Fisher forecast, showing that the current version of the analysis is satisfactorily close to the Fisher error. We regard this as a first step towards the realistic application of consistency relations to constrain primordial non-Gaussianity using observations.
\end{abstract}

\maketitle

\section{Introduction}\label{Sec:Intro}

One of the longest standing questions in modern cosmology concerns the mechanism that set the initial conditions for the classical evolution of matter in the Universe. In particular, one asks whether an initial inflationary period has taken place and, if so, what were the underlying physics. In this respect, a crucial aspect is to understand if the primordial fluctuations are consistent with those predicted by single-field inflation, or whether additional light degrees of freedom played a role. To answer this, one typically relies on data for fluctuations that are small enough to lie within the quasi-linear regime, whereupon perturbation theory provides reliable predictions. Such a theory can then be used to robustly extract the physical parameters. In contrast, modes at larger momentum (e.g., $k\gtrsim 0.2$~$h$/Mpc for large-scale structures at redshift zero) fall in the non-linear regime and are usually discarded due to modelling difficulties, despite being abundant and precisely measured.  It would be ideal to have a tool able to take advantage of this data.  

One possibility arises from the ``consistency relations," which are identities connecting the squeezed limit of $(N+1)$-point correlators (i.e., those with one vanishing momentum leg) and $N$-point correlators. Importantly, they follow solely from symmetry arguments and thus remain valid in the non-linear regime, where conventional perturbation theory fails. After being first pointed out by Maldacena~\cite{Maldacena:2002vr}, intensive work has been done in discovering additional relations arising from new symmetries as well as in clarifying the assumptions underlying their formulation~\cite{Creminelli:2004yq,Cheung:2007sv,Tanaka:2011aj,Creminelli:2012ed,Hinterbichler:2012nm,Assassi:2012zq,Kehagias:2012pd,Pajer:2013ana,Hinterbichler:2013dpa,Goldberger:2013rsa,Baldauf:2015xfa,Bravo:2017gct,Hui:2018cag}. In the context of large-scale structure, consistency relations can be traced back to the Galilean invariance of the evolution equations, as first identified by Kehagias \& Riotto~\cite{Kehagias:2013yd} and Peloso \& Pietroni~\cite{Peloso:2013zw}, and further investigated in~\cite{Valageas:2013zda,Creminelli:2013poa,Valageas:2016hhr,Marinucci:2019wdb,Marinucci:2020weg,Creminelli:2013mca}. 

Crucially, the consistency relations for large-scale structure are modified depending on whether the initial conditions for the matter fluctuations are Gaussian, as verified non-perturbatively in~\cite{Esposito:2019jkb}.  More precisely, if the initial conditions correspond to those of single-field slow-roll inflation, we expect
$B({\bm q}, {\bm k}) /P(q)$ to have no inverse-powers-of-$q$ pole.
Here, $B$ and $P$ are the bispectrum and power spectrum respectively, and ${\bm q}, {\bm k}$ are the momenta with the understanding that $q \ll k$ (the squeezed limit). 
Moreover, the symmetries from which the consistency relations originate characterize not only the evolution of matter distribution, but also that of astrophysically complex objects, such as galaxies (which can merge, form, annihilate, and so on)~\cite{Kehagias:2013rpa,Creminelli:2013mca, Simonovic:2014yna,Kehagias:2015tda}. Finally, consistency relations are expected to hold true in redshift-space~\cite{Creminelli:2013mca, Creminelli:2013poa,Creminelli:2013nua}, making them amenable for application to galaxy surveys, which include strong anisotropy around the line-of-sight direction due to redshift-space distortions.

In this work, we take an important step towards the practical implementation of consistency relations as a tool to obtain non-perturbative information on the initial conditions of the Universe, by studying the squeezed limit of the matter bispectrum. Using a suite of $N$-body simulations of matter in real-space, we show that the consistency relations can be used to precisely measure the local primordial non-Gaussianity parameter, $f_{\rm NL}$. The idea is that if the initial conditions were {\it not} that of single-field slow-roll inflation (the curvaton \cite{Lyth:2001nq,Zaldarriaga:2003my} model being an example, characterized by a non-vanishing $f_{\rm NL}$), the squeezed $B/P$ is no longer protected from having poles in the soft momentum $q$. \revision{In what follows, we consider only local-type non-Gaussianities which correspond to integer order poles as we will show in Section~\ref{sec: sq-bk}. Alternative inflation models, such as quasi-single field inflation \cite{Chen:2009zp, Baumann:2011nk, Assassi:2012zq, Noumi:2012vr}, can lead to non-integer order poles or oscillatory behavior in the squeezed $B/P.$ We leave a discussion of such models to a future work.}

We first derive the non-perturbative relation between the squeezed limit of the bispectrum and $f_{\rm NL}$ in Section~\ref{sec: sq-bk}, correcting and clarifying some results present in the literature. Section~\ref{sec: method} then highlights and addresses a number of subtleties regarding the procedure of measuring and analyzing the bispectrum in this context, the union of which allows for robust constraints on primordial non-Gaussianity, as presented in Section~\ref{sec: results}. Finally, we conclude and discuss the ingredients necessary for a future, realistic application of consistency relations to large-scale structure data in Section~\ref{sec: conclusion}.

\revision{The appendices give useful theoretical results and methodology validation. Appendix~\ref{appen: sq-bk-loop} includes tree-level and one-loop standard Eulerian perturbation theory calculations of the squeezed bispectrum in the presence of local primordial non-Gaussianity. Appendix~\ref{app:validation} validates our bispectrum parameterization. We derive our likelihood and Fisher matrix in Appendix~\ref{app:likelihood} and present a quasi-optimal weighting scheme in Appendix~\ref{appen: weights}. In Appendix~\ref{app:posterior} we show the full posterior of our main analysis.}

\vspace{1em}

\noindent\emph{Conventions:} Throughout this paper we work in natural units, $c=1$. Our fiducial cosmological parameters are $\Omega_{m,0} = 0.25$ (of which $\Omega_b = 0.04$), $\Omega_\Lambda = 0.75$, $h = 0.7$, $n_s = 1$, and $\sigma_8 = 0.8$.

\section{Squeezed bispectrum in the presence of local primordial non-Gaussianity}\label{sec: sq-bk}

\subsection{Derivation}

Let us begin by deriving the squeezed limit of the bispectrum in the presence of local primordial non-Gaussianity. We will work non-perturbatively in gravitational evolution, and follow a similar approach to the derivation of scale-dependent bias found in~\cite{Desjacques:2016bnm}. Additionally, we assume a single redshift, $z$, and generally suppress this argument for clarity. In the presence of a background (soft) potential, $\Phi_L$, the locally measured small-scale power spectrum is modulated, taking the form,\footnote{In our notation, the power spectrum and bispectrum are defined as $\langle \delta_{\vq} \delta_{\vk} \rangle  \equiv V \delta^{\rm K}_{\vq + \vk} P(q)$ and $\langle \delta_{\vq} \delta_{\vk} \delta_{\vp} \rangle \equiv V \delta^{\rm K}_{\vq + \vk + \vp} B(\vq,\vk)$ respectively, with $V$ the volume and $\delta^{\rm K}$ the Kronecker delta.}
\begin{align}
    \begin{split}
        P(k|\Phi_L) = P(k) + \sum_{\vq^\prime} \frac{\partial P(k)}{\partial\Phi_{L,\vq^\prime}}\Phi_{L,\vq^\prime} + \dots \,,
    \end{split}
\end{align}
where we are assuming that $q^\prime\equiv |\vq'| \ll k$, such that we can treat the long mode as a background, in presence of which the power spectrum is evaluated. This induces a coupling between the hard mode power spectrum, $P(k)$, and the soft mode density, $\delta_\vq$. The bispectrum is then obtained from the $\av{\delta_{\bm q} P(k|\Phi_L)}$ correlator, and is given by
\begin{align}\label{eq: bspec_def}
    \begin{split}
        B(\vq,\vk) &= V \sum_{\vq^\prime} \frac{\partial P(k)}{\partial \Phi_{L,\vq^\prime}} \av{\delta_\vq \Phi_{L,\vq^\prime}} + \dots \\
        &= V \frac{\partial P(k)}{\partial \Phi_{L,-\vq}}\frac{P(q)}{\alpha(q)}+\dots \,,
    \end{split}
\end{align}
where $V$ is the volume. Here $P(q)$ is soft and thus in the linear regime. This result uses the Poisson equation to relate the long-wavelength density and potential, $\delta_{\vq}=\alpha(q)\Phi_{L,\vq}$, with
\begin{align} \label{eq: alpha-q-def}
	\begin{split}
		\alpha(q) \equiv \frac{2}{3}\frac{q^2T(q)D_{\rm md}(z)}{\Omega_{m,0}H_0^2} \,,
	\end{split}
\end{align}
where $\Omega_{m,0}$ and $H_0$ are the matter density and expansion rate today, $T(q)$ is the transfer function (normalized to unity for $q \to 0$), and $D_{\rm md}(z)$ is the growth rate, normalized to $a(z)=1/(1+z)$ in the matter-dominated era.\footnote{Note that this differs from the normalization used in many cosmological codes, e.g., \textsc{class}.} We are interested in the form of Eq.~\eqref{eq: bspec_def} in the presence of local primordial non-Gaussianity, which can be parametrized by a primordial gravitational potential at sub-horizon scales given by~\citep[e.g.,][]{Komatsu:2001rj,Scoccimarro:2011pz},
\begin{align}
    \begin{split}
       \Phi(\bm x) = \phi(\bm x) + f_{\rm NL}\big(\phi^2(\bm x) - \langle \phi^2\rangle  \big) \,,
    \end{split}
\end{align}
where $\phi(\bm x)$ is a Gaussian random field with expectation value $\langle\phi\rangle$.

In the separate Universe picture, the soft mode acts as a background rescaling, equivalent to varying the locally measured clustering amplitude, $\sigma_8$, i.e.
\begin{align}\label{eq: sq-Bk-tmp1}
    \begin{split}
        B(\vq,\vk) = V\frac{\partial P(k)}{\partial\log\sigma_8}\frac{\partial\log\sigma_8}{\partial \Phi_{L,-\vq}}\frac{P(q)}{\alpha(q)}+\dots \,.
    \end{split}
\end{align}
To compute the potential derivative, we utilize the definition of $\sigma_8$ and $f_{\rm NL}$ \citep[e.g.,][]{Slosar:2008hx,Chiang:2017jnm,Giri:2022nzt}:
\begin{align}\label{eq: s8-fNL}
	\begin{split}
		\sigma_8(\Phi_L) = \sigma_8(0)\bigg[1 + \frac{2f_{\rm NL}}{V} \sum_{\vq^\prime} \Phi_{L,\vq^\prime} + \mathcal{O}\big(f_{\rm NL}^2\big)\bigg] \,,
	\end{split}
\end{align}
where we have noted that $\Phi_L(\bm x)$ is a slowly varying field, thus we can neglect its spatial dependence.
Combining Eqs.~\eqref{eq: alpha-q-def}, \eqref{eq: sq-Bk-tmp1}, and \eqref{eq: s8-fNL}, we find the leading term in the squeezed bispectrum to be,
\begin{align} \label{eq: sq-Bk-fNL}
	\begin{split}
		B(\vq,\vk) = \frac{6f_{\rm NL}\Omega_{m,0}H_0^2}{D_{\rm md}(z)} \, \frac{\partial P(k)}{\partial\log\sigma_8^2} \, \frac{P(q)}{q^2 T(q)} +\mathcal{O}\big(f_{\rm NL}^2\big) \,.
	\end{split}
\end{align} 
The result above differs from some former works in important ways. First, it may appear to differ from \citep[e.g.,][]{Simonovic:2014yna,Baldauf:2010vn}, which present a rigorous tree-level definition and find a $k$-dependence of the form $P(k)$ rather than $\partial P(k)/\partial\log\sigma_8^2$. This discrepancy is resolved by noting that, at first (tree-level) order, $P(k) \propto \sigma_8^2$, and thus the two coincide. More importantly, a non-perturbative derivation of the squeezed bispectrum is also presented in~\cite{Peloso:2013zw}, but neither the logarithmic derivative nor the $D_{\rm md}(z)$ factor appear in the final result. The presence of these factors is also verified by a direct one-loop calculation, as we report in Appendix~\ref{appen: sq-bk-loop}. In what follows, we further confirm their necessity in order to recover the correct $f_{\rm NL}$ when the latter is extracted from simulations including primordial non-Gaussianity.

\subsection{Modeling of the bispectrum} \label{sec:modeling}

To extract the primordial signal given in Eq.~\eqref{eq: sq-Bk-fNL}, we require a model for the squeezed bispectrum that also encodes the contributions coming from the non-primordial gravitational evolution. Beyond perturbative scales, the functional form of these is not well understood. Nonetheless, consistency relations tell us that the dependence on the soft mode is known, and that these late time contributions to the squeezed bispectrum scale with positive powers of the soft mode, $q$. In light of this, we write the squeezed bispectrum in the following form,

\begin{align} \label{eq: B-param}       
    \begin{split}
        B(q,k,\theta) ={}& a_{-2}(k) \frac{P(q)}{q^2 T(q)} + a_{-1}(k,\theta)\frac{P(q)}{q \, T(q)} \\
        & + \sum_{n=0}^\infty a_n(k,\theta) q^n P(q) \,,
    \end{split}
\end{align}
where we choose a parametrization in terms of the magnitude of the softest momentum, $q$, the magnitude of one of the hard modes, $k$, and the angle between them, $\theta$. Due to the consistency relations, $a_{-2}$ and $a_{-1}$ can only be sourced by primordial physics beyond single-field slow-roll inflation.\footnote{This is no longer true for unequal time correlators, which are expected to feature a (computable) $1/q^2$ term even in absence of primordial non-Gaussianity~\cite{Horn:2014rta}.} \revision{Our parametrization includes only integer order poles since we are assuming local primordial non-Gaussianity.} For the range of soft momenta we consider here the transfer function cannot be represented as a simple power series, thus we must include it explicitly. Inspired by perturbation theory, we only include it in the first two terms, since the remaining pieces are dominated by the bispectrum due to gravitational evolution, rather than by the primordial one, see Eq.~\eqref{eq: sq-Bk-fNL}. We will validate this parametrization in Section~\ref{sec: results}.

\begin{figure}
    \centering
    \includegraphics[width=\linewidth]{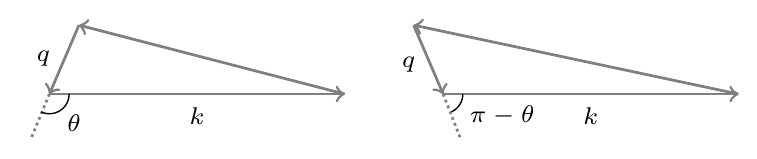}
            
    \caption{Triangles related by the parity transformation reported in Eq.~\eqref{eq:parity}. When all pairs of this sort are added together, the odd terms in the expansion of the bispectrum vanish.} \label{fig:triangles}
\end{figure}

A comparison with Eq.~\eqref{eq: sq-Bk-fNL} tells us that, at leading order in $f_{\rm NL}$, local primordial non-Gaussianity leads to a non-zero $a_{-2}$ coefficient,
\begin{align}\label{eq: am2-fNL}
    a_{-2}(k) = \frac{6 f_{\rm NL} \Omega_{m,0} H_0^2}{D_{\rm md}(z)}P(k)\frac{\partial \log {P(k)}}{\partial \log\sigma_8^2} \,,
\end{align}
while the other non-perturbative coefficients are a priori unknown.\footnote{A prediction for $a_0$ can actually be made, as shown and verified in~\cite{Nishimichi:2014jna}. Nonetheless, it relies on the non-linear observable being mass fluctuations, as opposed to, for example, galaxy ones. Given the scope of our broader program, we will remain agnostic about the nature of all non-perturbative coefficients, except for $a_{-2}$.} 
Nonetheless, one can still use invariance under parity to argue that~\cite{Lewis:2011au}
\begin{align} \label{eq:parity}
    a_n(k,\pi-\theta) = {(-1)}^n a_n(k,\theta) \,.
\end{align}
This means that, if one properly averages the bispectrum over all possible relative angles, the contributions to the odd coefficients coming from triangles related by the above parity transformation cancel (see Figure~\ref{fig:triangles}). This reduces the set of unknown parameters to the even coefficients.  We will return to this aspect in Section~\ref{subsec:measurement_procedure} and Appendix~\ref{app:validation}.

Our plan of action is the following: use $N$-body simulations to measure the $a_n$ coefficients (properly averaged over angles and hard momenta, as explained in Section~\ref{subsec:measurement_procedure}), then extract $f_{\rm NL}$ from $a_{-2}$ using Eq.~\eqref{eq: am2-fNL}.

\section{Methodology}\label{sec: method}

To validate our measurements of $f_{\rm NL}$, we use a suite of $N$-body simulations consisting of 40 realizations with Gaussian initial conditions and 12 realizations with local primordial non-Gaussianity ($f_{\rm NL}=100$). The simulations were run with $1280^3$ particles in a comoving cubic volume of $L^3 = (2.4$~${\rm Gpc}/h)^3$. The cosmological parameters are $\Omega_{m,0}=0.25$ (of which $\Omega_b=0.04$), $\Omega_\Lambda=0.75$, $h=0.7$, $n_s=1$, and $\sigma_8=0.8.$ We analyze the $z=0$ and $z=0.97$ snapshots. For further details on the simulations, we refer the reader to~\cite{Scoccimarro:2011pz}.

\subsection{Measuring the power spectrum and bispectrum} \label{subsec:measurement_procedure}

To obtain an estimate of $f_{\rm NL}$ using the consistency relations, we will require both the matter power spectrum and bispectrum. These are measured using fast Fourier transforms, first assigning the simulation particles to a $1024^3$ mesh using Triangular Shaped Cloud interpolation. The soft momentum is collected in bins of width $2k_f$, with $k_f\equiv 2\pi/L\simeq 2.6 \times 10^{-3}$~$h$/Mpc being the fundamental mode, starting from modes of magnitude $q>k_f$ up to $q_{\rm max}=25k_f \simeq 0.065$~$h$/Mpc---i.e. 12 soft momentum bins. The hard modes entering the bispectrum will, instead, be taken within a range that is variable, depending on our analysis choices (as discussed below).

Throughout our analysis we pay particular attention to avoid any artifacts induced by binning and the discrete nature of the momenta. Such effects are more prominent for the softest momenta which, at the same time, are the most sensitive to the signal arising from primordial non-Gaussianity.
For this reason, as far as the power spectrum and its related quantities are concerned, we refrain from any modeling (i.e., using linear theory) and directly measure the following binned quantities,
\begin{align} \label{eq:Mhat}
    \hat{M}_{i,n} \equiv 
    \begin{cases}
        \big\langle q^n\delta_{\bm q}\delta_{-\bm q}/T(q) \big\rangle_i \,, & \text{if}\ n<0 \\
        \big\langle q^n\delta_{\bm q}\delta_{-\bm q} \big\rangle_i \,, & \text{otherwise}
    \end{cases} \,,
\end{align}
where by $\langle \, \cdots \rangle_i$ we represent the average over $\bm{q}$ modes whose magnitude falls within the $i$-th bin. By directly measuring $\hat{M}_{i,n}$ from the simulations, we also take advantage of the sample variance cancellation arising from the fact that bispectrum and power spectrum are measured from the same density field.

For the bispectrum, we measure the following quantities
\begin{align} \label{eq:Bhat}
    \hat{B}_i \equiv \big\langle {\rm Re}\, \delta_{\bm q} \delta_{\bm k} \delta_{-\bm q-\bm k} \big\rangle_i \,, \quad \text{ with } k \in [k_{\rm min}, k_{\rm max}] \,,
\end{align}
where $k_{\rm min}$ and $k_{\rm max}$ define the hard momentum scales of our analysis. We sum over all $\bm{k}$'s with magnitude between these values.
More precisely, if the relative angle between $\bm{q}$ and $\bm{k}$ is allowed to cover the full range between $0$ and $2\pi$, then the hard modes contributing to the bispectrum in the soft momenta bin $q_i$ effectively range from $k_{\rm min} - q_i$ to $k_{\rm max} + q_i$, and hence the hard momenta range varies with $q$.  In order to avoid finite resolution effects from the simulation or our particle assignment scheme, we impose a maximum momentum in our analysis of $K = 270k_f \simeq 0.71$~$h$/Mpc. This means that we must always make sure that $k_{\rm max} + q_{\rm max} \leq K$. Once this is done, we are guaranteed that the average in Eq.~\eqref{eq:Bhat} is done over all possible relative angles, $\theta$, hence ensuring that the odd coefficients in the expansion in Eq.~\eqref{eq: B-param} average to zero, as discussed below Eq.~\eqref{eq:parity}.
Moreover, we simply average over all $k$'s between $k_{\rm min}$ and $k_{\rm max}$ with uniform weight, instead of treating different $k$-bins separately. This way we can remain agnostic about the $k$-dependence of all coefficients in Eq.~\eqref{eq: B-param} except for $a_{-2}(k)$, otherwise we would need to introduce a new set of (unknown) coefficients for each $k$-bin.\footnote{Alternatively, one could average the bispectrum over different $k$-modes using an optimal weighting scheme. 
We tested several quasi-optimal weighting strategies (derived in Appendix~\ref{appen: weights}) and found negligible improvement on our constraints, thus we defer a discussion of weighting to a future work. The Fisher forecast presented in Section~\ref{subsec: forecasts} suggests that our analysis is fairly close to optimal even without including $k$-dependent weights.}

In light of the discussion above, our theory model for the bispectrum is given by
\begin{align} \label{eq: theory_model}
    \hat{\bm{B}} = \sum_{n=-2,0,2}\bar{a}_n \hat{\bm M}_{n} \,,
\end{align}
where vectors run over the soft mode index, $i$, and the $\bar{a}_n$ coefficients correspond to the $a_n$'s in Eq.~\eqref{eq: B-param} averaged over angles and over hard modes with magnitude between $k_{\rm min}$ and $k_{\rm max}$. In Appendix~\ref{app:validation} we show that repeating the analysis with the inclusion of the odd coefficients ($\bar{a}_{-1}, \bar{a}_1$) leads to values consistent with zero. This confirms that the average over triangles is done properly, and it is therefore sensible to only consider the subset of even coefficients. We also show that $\bar{a}_4$ is consistent with zero so that we can truncate the series at $\bar{a}_2.$ 

Finally, in order to extract $f_{\rm NL}$ from $\bar{a}_{-2}$, we measure $\bar{F} \equiv\big \langle \delta_{\bm k}\delta_{-\bm k} \, \partial \log P(k)/\partial \log \sigma_8^2 \big\rangle$, averaged over $k \in [k_{\rm min}, k_{\rm max}]$, using the same bins as the bispectrum. A couple of subtleties arise when estimating this quantity. First, in principle, $P(k)$ should be computed from a theoretical model as it is an average over realizations; however, since we do not a priori know $P(k)$ in the non-linear regime, we instead measure it from the simulations. This does not significantly impact our results since there is little variance in $P(k)$ at high-$k$.  Second, $\partial \log P(k)/\partial \log \sigma_8^2$ should be determined from simulations, for example, via a separate Universe prescription; however, this is a computationally expensive and time consuming process. For our fiducial analysis it suffices to estimate the function using \textsc{halofit}~\cite{Takahashi:2012em}. This represents the main source of theoretical uncertainty when constraining $f_{\rm NL}$ from consistency relations using the real-space matter distribution.\footnote{To elucidate the importance of this theoretical error, we have repeated the analysis instead taking the $\partial \log P(k) / \partial \log \sigma_8^2$ derivative from the \textsc{Quijote} simulations \cite{Villaescusa-Navarro:2019bje} at their fiducial cosmology. We find that the final value of $f_{\rm NL}$ changes by less than $5\%$, which is smaller than our statistical uncertainty (see Section~\ref{sec: results}).} 
For a measured $\bar{F}$ and $\bar{a}_{-2}$ (averaged over the same hard momenta), we can compute $f_{\rm NL}$ using the relationship
\begin{align}
    f_{\rm NL}=\frac{\bar{a}_{-2}}{\bar F}\frac{D_{\rm md}(z)}{6\Omega_{m,0}H_0^2} \,.
\end{align}

\pagebreak

\subsection{Likelihood and covariance}\label{Sec: likelihood}

To formulate our likelihood for the non-perturbative coefficients, $\bar{a}_n$, we must account for the stochastic nature of both the measured bispectrum $\hat{\bm B}$ and the theory model, the latter of which depends on the measured binned power spectra, $\hat{\bm M}_n$, which we also consider as a stochastic variable. As shown in Appendix~\ref{app:likelihood}, this can be taken into account by a likelihood that, for a single realization, $r$, is given by
\begin{align} \label{eq:likelihood}
    \mathcal{L} \propto \frac{1}{\sqrt{\det \bm{\mathcal{C}}(\bar{a}_n)}}\bigg[1+\frac{\delta\bm B(\bar{a}_n)\cdot\bm{\mathcal{C}}^{-1}(\bar{a}_n)\cdot\delta \bm B(\bar{a}_n)}{N_R-1}\bigg]^{-\frac{N_R}{2}} \,,
\end{align}
where $\delta \bm B(\bar{a}_n) \equiv \hat{\bm B}^r- \sum_n \bar{a}_n \hat{\bm M}_n^r$ is the difference between the measured bispectrum and our theory model (Eq.~\ref{eq: theory_model}), 
${\mathcal{C}}_{ij}(\bar{a}_n)\equiv \langle\delta B_i(\bar{a}_n) \delta B_j(\bar{a}_n)\rangle$ is the \emph{parameter dependent} covariance of $\delta \bm B$, and $N_R$ is the number of realizations used to estimate it. Note that we use a multivariate $t$-distribution instead of a Gaussian likelihood to account for the fact that our covariance is estimated from a finite number of realizations~\cite{Sellentin:2015waz} as discussed below.

Since a real analysis would assume a fiducial value of $f_{\rm NL}=0$ for the covariance and we only have twelve realizations with primordial non-Gaussianity, we always compute our covariance using the $N_R=40$ realizations with Gaussian initial conditions. Furthermore, to ease the computational burden associated with sampling a likelihood with a parameter dependent covariance, we evaluate the latter at a fixed set of fiducial $\bar{a}_n$'s. Evaluating the covariance at a fiducial set of parameters is also necessary so that we can use covariance from simulations without primordial non-Gaussianity when fitting measurements from simulations with $f_{\rm NL}=100$. To determine the fiducial set of coefficients we use an iterative procedure (similar to a Newton--Raphson algorithm) to maximize the likelihood for $\hat{\bm B}$ and $\hat{\bm M}_n$, which are the mean measurements from the set of simulations with Gaussian initial conditions.

Specifically, we first maximize the likelihood with a covariance evaluated at $\bar a_n = 0$. Then, we use the new coefficients to re-compute the covariance, which is used to re-maximize the likelihood, and so on. We stop the procedure once all parameters have converged with a fractional accuracy of $10^{-9}$, which takes around 10 iterations. We have checked that this procedure does not affect our results in an appreciable way.\footnote{Our fiducial results for the simulations with Gaussian initial conditions change from $f_{\rm NL}=-0.6 \pm12$ to $f_{\rm NL}=-0.1\pm12$ when including the parameter dependent covariance. } Once the set of fiducial coefficients has been determined, and thus the covariance, $\bm{\mathcal{C}}$, we sample from the parameter posterior distribution. For our analysis, we vary the $\bar{a}_n$ coefficients with uniform wide priors between  $-10^9$ and $10^9~({\rm Mpc}/h)^{3+n}$. The sampling is done using the \texttt{Multinest} package~\cite{Feroz:2008xx} with 500 live points and an evidence tolerance of 0.005. 

\begin{figure*}
\includegraphics[width=\linewidth]{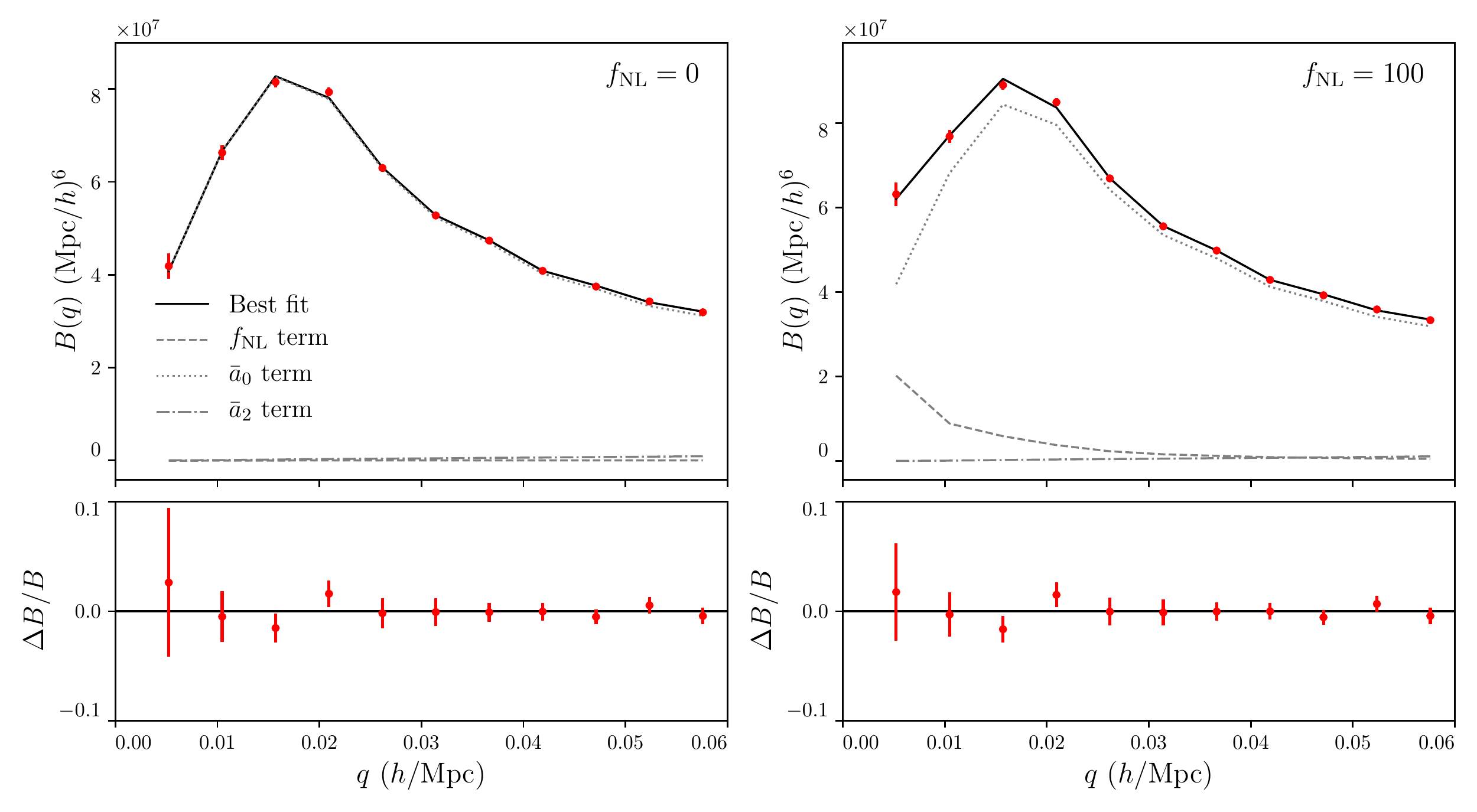}
\caption{\textbf{Top panel:} measured hard-mode-averaged squeezed bispectrum (red points) and maximum a posteriori theory prediction (black lines) for a single realization of the $z=0$ snapshots without (\textbf{left}) and with (\textbf{right}) local primordial non-Gaussianity assuming $q_{\rm max} = 23k_f \simeq 0.06~h/{\rm Mpc}$, $k_{\rm min}=  75k_f \simeq 0.2$~$h$/Mpc, and $k_{\rm max} = 231k_f\simeq 0.6$~$h$/Mpc. Gray lines indicate the individual contributions of each component of the theory model. In the presence of local primordial non-Gaussianity, there is a sizable contribution from the $1/q^2$ term in the squeezed bispectrum. The error bars are computed from the fiducial covariance described in Section~\ref{Sec: likelihood}, scaled to match the volume of a single realization. 
\textbf{Bottom panel:} fractional residuals with respect to the maximum a posteriori prediction. } \label{fig:bspec_measurement}
\end{figure*}
\begin{figure*}
\includegraphics[width=\textwidth]{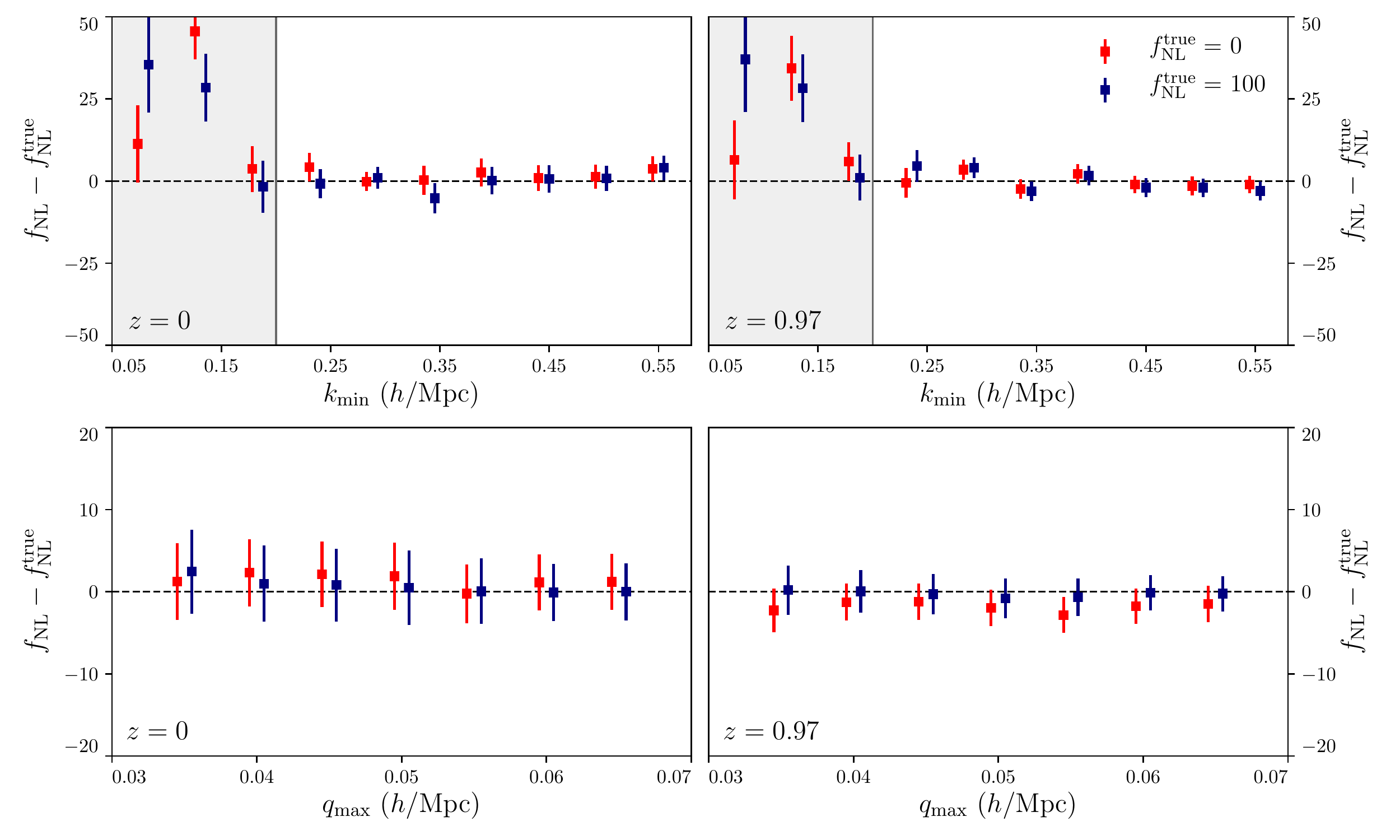}
\caption{\textbf{Top panel:} mean value for $f_{\rm NL}$ and its $68\%$ error, for different sets of hard modes at redshift $z=0$ (\textbf{left}) and $z=0.97$ (\textbf{right}), and for both Gaussian (\textbf{red}) and non-Gaussian (\textbf{blue}) initial conditions. The maximum soft momentum has been fixed to $q_{\rm max} = 23k_f \simeq 0.06$~$h$/Mpc. The hard modes have been collected in a thin bin of size $\Delta k = 2k_f \simeq 0.013$~$h$/Mpc, with varying $k_{\rm min}$. The shaded gray region denotes modes with $k_{\rm min}< 75 k_f \simeq 0.2$~$h/{\rm Mpc}$, which are excluded in the remainder of our analyses. In this regime, triangles are not sufficiently squeezed, thus our estimates of $f_{\rm NL}$ become biased.
\textbf{Bottom panel:} as above, but fixing the hard momentum range to be $k_{\rm min} = 75 k_f \simeq 0.2$~$h$/Mpc and $k_{\rm max} =231 k_f \simeq 0.6$~$h$/Mpc and varying the maximum soft momentum, $q_{\rm max}$. All the error bars refer to a volume corresponding to 12 realizations. These results confirm the expectation that consistency relations are only applicable when $q/k$ is sufficiently small, and that most of the signal is contained in the lowest $q$-modes.} \label{fig:kmin_dependence}
\end{figure*}
\begin{figure*}
\includegraphics[width=\textwidth]{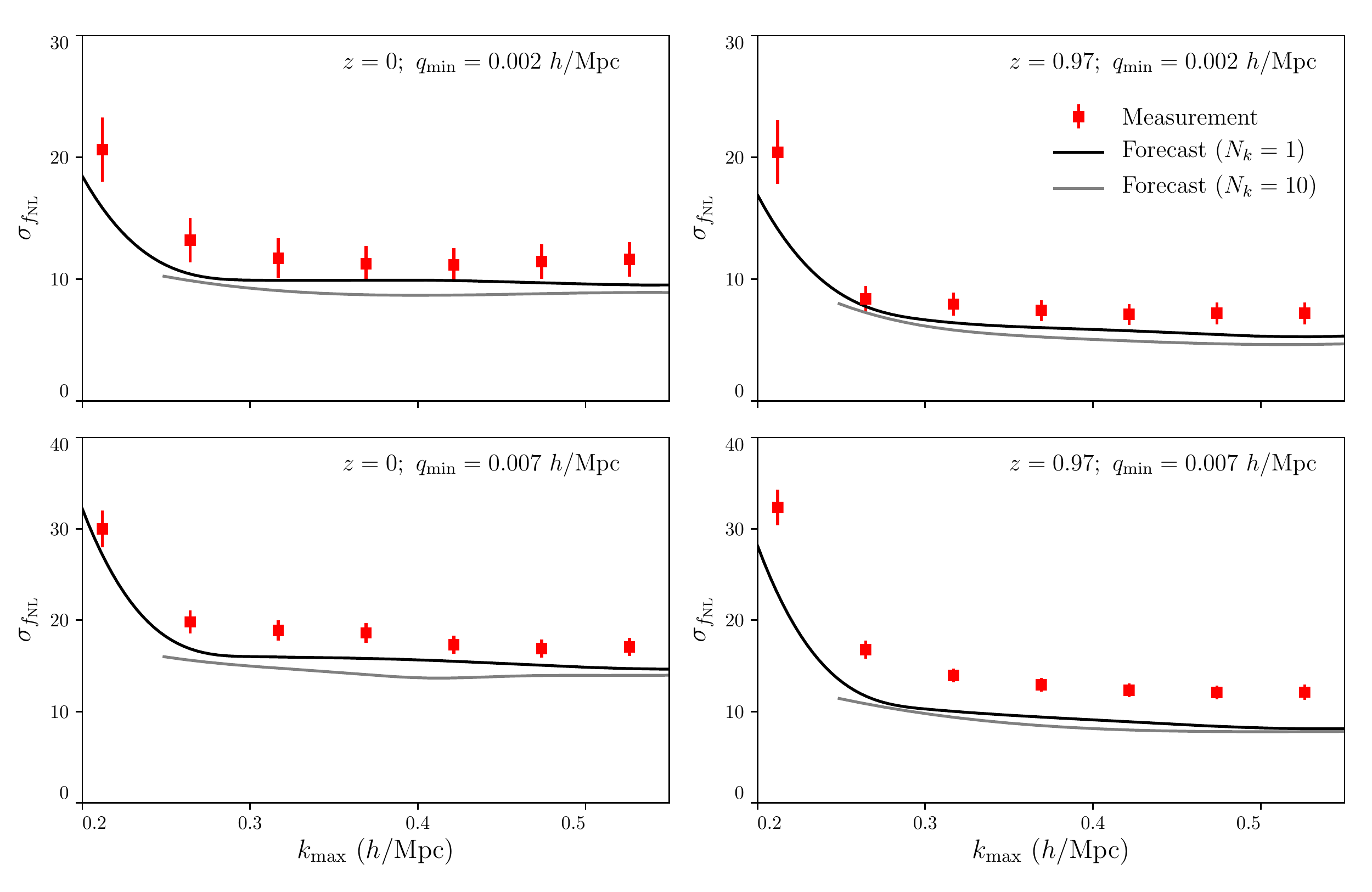}
\caption{Comparison of the uncertainty on $f_{\rm NL}$ from our analysis of 40 simulations with a volume of $(2.4 \text{ Gpc}/h)^3$ and $f_{\rm NL}=0$ (points) with forecasted errors for the same volume (lines). The error is shown as a function of $k_{\rm max}$ where $k_{\rm min}=75k_f\simeq 0.2~h/{\rm Mpc}$ has been fixed for all analyses. Constraints are shown with ({\bf top}) and without ({\bf bottom}) the softest momentum bin for redshift $z=0$ ({\bf left}) and $z=0.97$ ({\bf right}). The Fisher forecast is performed assuming all modes have been averaged in a single $k$-bin ({\bf black}), matching our analysis, as well as splitting the data into $10$ $k$-bins ({\bf grey}). The improvement from treating the $k$ bins separately is less than 10\%, which suggests that there is little to gain from using a more complex hard momentum weighting scheme and/or introducing different values of $\bar{a}_0$ and $\bar{a}_2$ for each $k$ bin. The error bars quickly saturate as a function of $k_{\rm max}$. }
\label{fig:fisher_forecasts}
\end{figure*}
\section{Results}\label{sec: results}

We first present a series of fiducial results carried out on measurements from a single realization of our simulations at two redshifts ($z=0$ and $z=0.97$) with soft momenta $0.005<q<0.06~h/{\rm Mpc}$ and hard momenta $0.2<k<0.6~h/{\rm Mpc}$. The full posteriors for this analysis can be found in Appendix~\ref{app:posterior}. We will later validate this choice of scale cuts, and investigate the corresponding sensitivity of our $f_{\rm NL}$ constraints. 

In Figure~\ref{fig:bspec_measurement} we plot the squeezed bispectrum and our maximum a posteriori model prediction for a single realization at redshift $z=0$. The error bars are computed from the fiducial covariance introduced in Section~\ref{Sec: likelihood}. As expected, in the presence of local primordial non-Gaussianity, there is a sizable contribution from the $1/q^2$ term in the squeezed bispectrum. Moreover, the $\bar{a}_2$ contribution becomes significant at scales $q\gtrsim0.035~h/{\rm Mpc}$, matching the conclusions of Appendix~\ref{app:validation}. For this particular realization and choice of scale cuts, we find $f_{\rm NL}=0\pm 12$ and $f_{\rm NL}=98 \pm 12$, for the measurement without and with local primordial non-Gaussianity, respectively. Our constraints improve considerably at redshift $z=0.97$ for which we find  $f_{\rm NL}=-1 \pm 8$ and $f_{\rm NL} = 97 \pm 8$. 

The improvement at higher redshift can be understood as follows. On the one hand, the primordial contribution to the squeezed bispectrum, Eq.~\eqref{eq: sq-Bk-fNL}, scales as $B \sim P^2 / D(z) \sim D^3(z)$ for growth factor $D(z)$. On the other hand, the bispectrum covariance scales as $C_B \sim P^3 \sim D^6(z)$. It follows that the signal-to-noise ratio has the scaling $B/\sqrt{C_B} \sim D^0(z)$, which is redshift-independent. Nonetheless, at higher redshift, gravitational non-Gaussianity is suppressed by a factor of $D(z)$, which (a) reduces the $\bar{a}_n$ signal for $n>0$ (which must be disentangled from $f_{\rm NL}$), and (b) reduces the non-Gaussian covariance. These effects increase the signal-to-noise with which primordial non-Gaussianity can be measured at high-redshift. This indicates a transfer of information on $f_{\rm NL}$ into higher-point functions at late times.

\subsection{Squeezed criterion}\label{subsec: scalecuts}

In this section we fit the squeezed bispectra for different ranges of minimum hard momenta, $k_{\rm min}$, and maximum soft momenta, $q_{\rm max}$, to address how squeezed the bispectrum must be in order for consistency relations to be applied reliably. Since our focus in this section is on method validation, all results use the mean bispectrum measurement obtained from 12 realizations with Gaussian initial conditions, and 12 realizations with local primordial non-Gaussianity. The covariance is, however, always evaluated using the 40 Gaussian simulations, but scaled to match the volume of 12 realizations.

In the top panel of Figure~\ref{fig:kmin_dependence} we show the mean value for $f_{\rm NL}$ and its $68\%$ confidence error, for both Gaussian and non-Gaussian initial conditions, as well as for redshifts $z=0$ and $z=0.97$. We do this for different values of $k_{\rm min}$, but fixing $k_{\rm max} = k_{\rm min} + \Delta k$, with $\Delta k = 5k_f \simeq 0.013$~$h$/Mpc. We fix the minimum and maximum soft momenta to be $q_{\rm min} = k_f \simeq 0.002$~$h$/Mpc and $q_{\rm max} = 23k_f \simeq 0.06$~$h$/Mpc. For sufficiently high $k_{\rm min}$, our fit closely matches the expected value of $f_{\rm NL}$. However, for $k_{\rm min} \lesssim 0.15$~$h$/Mpc we obtain biased results, a signal of the fact that the triangles are not sufficiently squeezed for the relation derived in Eq.~\eqref{eq: sq-Bk-fNL} to apply. The scale at which we recover unbiased results appears to be consistent between the two redshifts. In light of the above results, unless otherwise stated, for the remainder of the analyses we fix the minimum hard momentum to $k_{\rm min} \simeq 0.2~h/{\rm Mpc}$. 

Having determined a range of suitable hard momentum scale cuts, we now analyze the dependence of our results on the soft momenta scale cuts. To do this, we fix the maximum hard momentum to $k_{\rm max} \simeq 0.6$~$h$/Mpc and the minimum soft momentum again to $q_{\rm min} = k_f \simeq 0.002$~$h$/Mpc, but vary the maximum soft momentum, $q_{\rm max}$. The scale cuts are such that each successive value of $q_{\rm max}$ contains one more bin than the previous one. We report the results in the bottom panel of  Figure~\ref{fig:kmin_dependence}. As one can see, we recover the correct value of $f_{\rm NL}$ for both sets of initial conditions, essentially independently of the value chosen for $q_{\rm max}$. The error on $f_{\rm NL}$ also decreases with increasing maximum soft momentum, albeit mildly. Both observations are explained by the fact that most of the signal is contained in the first few $q$-bins. For this reason, from now on we fix $q_{\rm max} = 23 k_f \simeq 0.06$~$h$/Mpc.

\revision{In summary, our constraints on $f_{\rm NL}$ from the matter bispectrum are unbiased for scale cuts with $k_{\rm min}/q_{\rm max}\gtrsim 3$ at redshifts $z=0$ and $z=0.97$. Since most of the information is contained within the softest momentum bin, the more relevant scale is likely set by $k_{\rm min}$ and $q_{\rm min}$. For our analysis, the associated ratio is then $k_{\rm min}/q_{\rm min}\approx30$ (100) for $q_{\rm min}=0.007~h/{\rm Mpc}$ ($0.002~h/{\rm Mpc}$).} Further discussion on the parametrization of the bispectrum and on the scale cuts can be found in Appendix~\ref{app:validation}.

\subsection{Information content and Fisher forecasts}\label{subsec: forecasts}

Having demonstrated that the consistency relations can be used to extract unbiased constraints on $f_{\rm NL}$ in the non-linear regime, we now analyze the information content on $f_{\rm NL}$ in the consistency relations. Of particular interest is how our constraints on $f_{\rm NL}$ change as we vary $q_{\rm min}$ and $k_{\rm max}$, as well as how our results are limited by na\"ively averaging over all hard momenta with uniform weight. To this end, we fit the squeezed bispectra for varying maximum hard momenta, $k_{\rm max}$, and minimum soft momenta, $q_{\rm min}$, and additionally perform a series of analytic Fisher forecasts. The derivation of the Fisher matrix used in these forecasts is presented in Appendix~\ref{appen: fisher_matrix}.

In Figure~\ref{fig:fisher_forecasts} we plot the error bar on $f_{\rm NL}$ as a function of $k_{\rm max}$ for redshift $z=0$ and $z=0.97$. We show the mean and the 68\% confidence limit for the error on $f_{\rm NL}$ computed by fitting each of the 40 realizations with Gaussian initial conditions individually. All results assume a maximum soft momentum $q_{\rm max} \simeq 0.06~h/{\rm Mpc}$ and minimum hard momentum $k_{\rm min}\simeq 0.2~h/{\rm Mpc}.$ The bottom panels exclude the first soft momenta bin. Removing the softest momenta bin increases the error on $f_{\rm NL}$ from $12\pm1$ ($7\pm1$) to $17\pm1$ ($12\pm 1)$ at $k_{\rm max}=0.54~h/{\rm Mpc}$ and $z=0$ ($z=0.97)$, confirming that most of the information is contained within the softest momenta bin.

Turning our attention to the the small-scale information, we find that the error on $f_{\rm NL}$ quickly saturates with increasing $k_{\rm max}.$ This is due to the non-Gaussian contributions to the covariance and is consistent with recent results analyzing local primordial non-Gaussianity in the non-linear regime \cite{Coulton:2022rir}. Thus, even though the consistency relations can be used to constrain $f_{\rm NL}$ at scales well beyond $0.35~h/{\rm Mpc},$ we find essentially no benefit when including those modes. The situation may be less dire at high redshifts as was found in \cite{Floss:2022wkq}. In particular, when we increase $k_{\rm max}$ from $121k_f\simeq0.32~h/{\rm Mpc}$ to $221k_f\simeq0.58~h/{\rm Mpc}$ we find that $\sigma_{f_{\rm NL}}$ remains essentially constant at $z=0$ ($11.7\pm 1.6$ to $11.6\pm 1.4$), but decreases by 10\% at $z=0.97$ ($8.0\pm 1.0$ to $7.2\pm0.9$). \revision{Given that the non-Gaussian contributions to the covariance are particularly important for local shapes, even at high redshifts \cite{Floss:2022wkq, Biagetti:2021tua}, the gains at higher redshift may be more modest than those found for other types of non-Gaussianities. }

To ensure that the lack of small-scale information is not an artifact of our sub-optimal weighting scheme (i.e., summing all $k$-modes with uniform weights), we perform a series of Fisher forecasts. The black curves show the forecasts assuming that all hard momenta are averaged into a single $k$ bin, as is the procedure used in our analysis. We find that our errors are within 20\% of the Fisher error for all scale cuts, with the exception of the $z=0.97$ snapshot with the first bin excluded, for which our error bars are 50\% larger. Note that for simplicity the Fisher forecasts assume $\bar{a}_2=0$, which leads to an underestimate of $\sigma_{f_{\rm NL}}$ due to the degeneracy between $\bar{a}_2$ and $f_{\rm NL}$, as discussed in Appendix~\ref{app:validation}.\footnote{A more fair comparison between our measurement error and the Fisher error on $f_{\rm NL}$ can be made by fixing $\bar{a}_2=0$ and choosing $q_{\rm max}\leq 0.035~h/{\rm Mpc}$ as was verified in Appendix~\ref{appen: fisher_matrix}. In this case we find that our measured error on $f_{\rm NL}$ is 7\% (14\%) larger than the Fisher error for $N_k=1$ $(N_k=10)$ at $z=0$ assuming $q_{\rm min}=0.002~h/{\rm Mpc}$, $q_{\rm max}=0.025~h/{\rm Mpc}$, $k_{\rm min}=76k_f$, and $k_{\rm max}=231k_f.$} 

We also perform the Fisher forecast assuming the hard momenta between $k_{\rm min}$ and $k_{\rm max}$ are subdivided into $N_k=10$ linearly spaced hard momenta bins. This is equivalent to the improvement that one would expect if we accounted for the different values of $\bar{a}_0$ over 10 bins, as opposed to simply averaging over all modes. Interestingly, there is only a very modest $(<10\%)$ improvement when treating the hard momenta bins separately, hence there is likely little to gain by implementing an optimal weighting scheme. We found that the error on $f_{\rm NL}$ does start to decrease if we consider $N_k> 20$; however, given that the covariance is estimated from only 40 realizations, we do not include those results and defer this study to a future work in which we have a more robust covariance.

Finally, using our forecasting pipeline, we present an \emph{extremely idealistic} forecast for the error on local $f_{\rm NL}$ that we can hope to achieve from the Dark Energy Spectroscopic Instrument (DESI) \cite{DESI:2016fyo} using our method. Assuming DESI will measure a combined survey volume of $170 \ {\rm Gpc}^3\simeq 58.4 ~({\rm Gpc}/h)^3$ (Table 3 of \cite{DESI:2013agm}) and neglecting the (likely relevant) complications due to galaxy bias and redshift-space distortions, we find $\sigma_{f_{\rm NL}}\simeq4.9$ and $3.3$ using the covariance at redshift 0 and 0.97, respectively, and scale cuts $q_{\rm min}\simeq 0.002, \ q_{\rm max}\simeq0.06, \ k_{\rm min}\simeq0.2, \ k_{\rm max}\simeq0.3~{h/{\rm Mpc}}$. Although these forecasts are highly optimistic, combining our analysis with existing techniques to constrain $f_{\rm NL}$ from galaxy clustering, particularly those that use the scale-dependent bias, may considerably improve our ability to constrain local primordial non-Gaussianity.

\section{Conclusions}\label{sec: conclusion}

Due to its ability to differentiate between single-field and multi-field inflation, measuring local primordial non-Gaussianity is a key target for current and upcoming cosmological surveys. Here we present a method to measure local $f_{\rm NL}$ from squeezed configurations of the real-space matter bispectrum using the consistency relations. Crucially, our analysis is non-perturbative, allowing us to reliably use data deep into the non-linear regime. To our knowledge, this is the first result using the consistency relations to directly measure local $f_{\rm NL}.$ Our main results are summarized as follows: 

\begin{itemize}
    \item Using a suite of $N$-body simulations with local $f_{\rm NL}=100$ and with covariance scaled to match a volume of $(2.4 \ {\rm Mpc}/h)^3$, we measure $f_{\rm NL} = 98 \pm 12$ at redshift $z=0$ and $f_{\rm NL} = 97 \pm 8$ at $z=0.97$ from the real-space matter bispectrum using the consistency relations.
    \item Due to the non-Gaussian bispectrum covariance at late times in the short-scale regime, our error on $f_{\rm NL}$ is relatively insensitive to the choice of maximum hard momenta, $k_{\rm max}$. In particular, we find little improvement when including modes beyond $k\gtrsim 0.35~h/{\rm Mpc}$ at redshifts 0 and 0.97, even though our model has been verified up to $k_{\rm max}\simeq 0.7~h/{\rm Mpc}.$
    \item Constraints on $f_{\rm NL}$ from the consistency relations are extremely sensitive to the amount of large-scale information. The error on $f_{\rm NL}$ increases by 50\% if we increase $q_{\rm min}$ from $0.002~{h/\rm Mpc}$ to $0.007~{h/\rm Mpc}$ at redshifts $z=0$ and $z=0.97.$
\end{itemize}

In light of these findings, we conclude that even though the consistency relations provide a robust mechanism to constrain local $f_{\rm NL}$ using measurements in the non-linear regime, the utility of this information is severely limited by the non-Gaussian covariance at small scales. We expect this effect to be less severe at high redshifts, although a more dedicated analysis is needed to verify this. Even if the consistency relations alone are unable to place competitive constraints on local $f_{\rm NL}$, they provide a strong test of fundamental physics with minimal modeling assumptions and applying an analysis like the one presented in this work to observational data would be highly informative.

Given that the approach presented here used the real-space matter distribution, it would most easily be generalized to galaxy or CMB lensing surveys (which measure only the matter field), albeit with loss of radial information. Furthermore, recent and upcoming lensing surveys such as the Dark Energy Survey \cite{DES:2005dhi}, the Legacy Survey of Space and Time \cite{LSSTScience:2009jmu}, Atacama Cosmology Telescope \cite{Kosowsky:2003smi}, Simons Observatory \cite{SimonsObservatory:2018koc}, and CMB-S4 \cite{CMB-S4:2016ple} will yield precise measurements of a wealth of non-linear modes, making them particularly useful for consistency relation-based methods.

On the other hand, since the consistency relations are expected to hold for biased tracers in redshift-space \cite{Kehagias:2013rpa,Creminelli:2013mca, Creminelli:2013poa,Creminelli:2013nua, Simonovic:2014yna,Kehagias:2015tda}, one could extend our method to the bispectrum of photometric galaxy surveys. The main ingredient necessary to apply our method to 2D projected galaxy clustering measurements is a treatment of galaxy bias. This may be non-trivial given that the hard momenta in our analysis are in the non-linear regime. Moreover, the relationship between the measured coefficient $\bar{a}_{-2}$ and $f_{\rm NL}$ (Eq.~\ref{eq: am2-fNL}) will need to be modified to account for the scale-dependent bias and the response of the non-linear galaxy power spectrum to a long-wavelength density perturbation, both of which could have significant theoretical uncertainties depending on the galaxy population \cite{Barreira:2022sey}.

One could also expand the method presented in this work to 3D galaxy clustering measurements from spectroscopic galaxy surveys, such as BOSS \cite{Reid:2015gra}, DESI \cite{DESI:2016fyo}, Euclid \cite{Amendola:2016saw}, and SPHEREx \cite{Dore:2014cca}. In addition to accounting for galaxy bias, this would require modeling redshift-space distortions and accounting for stochastic effects, i.e. shot noise. Such an analysis would complement and extend existing analyses which measured local $f_{\rm NL}$ from the redshift-space galaxy bispectrum in the perturbative regime \cite{Cabass:2022ymb, DAmico:2022gki} and reveal if the consistency relations contain information beyond the scale-dependent bias. 

The consistency relations relate correlators of all orders, hence there is likely information from the consistency relations that we are not capturing in our analysis. This could be probed by considering higher-order correlators such as the trispectrum, but estimating such quantities and their covariances is both computationally and theoretically challenging. An alternative would be to constrain primordial non-Gaussianity at the field level which, by design, incorporates information from higher-order correlation functions \cite{Giri:2022nzt,Andrews:2022nvv}. A comparison with field level approaches would be extremely useful to quantify the information content of the consistency relations. 

\revision{Finally, one could generalize the method presented in this work to constrain other types of non-Gaussianities that violate the consistency relations. The most direct extension would be to consider non-integer poles as are predicted in quasi-single field inflation scenarios \cite{Chen:2009zp, Baumann:2011nk, Assassi:2012zq, Noumi:2012vr}. This would require modifying our bispectrum parametrization to include such poles and their appropriate coefficients.}

\acknowledgments

We thank Kazuyuki Akitsu for sharing code used to compute the bispectrum. We are grateful to Matteo Biagetti, Will Coulton, Mikhail Ivanov, Kendrick Smith, Marko Simonovi\'c, and Beatriz Tucci for insightful discussions. \revision{We thank the referee for their helpful comments and suggestions.} For most of the development of this work A.E. has been a Roger Dashen Member at the Institute for Advanced Study, whose work was also supported by the U.S. Department of Energy, Office of Science, Office of High Energy Physics under Award No. DE-SC0009988. 
O.H.E.P. is a Junior Fellow of the Simons Society of Fellows and thanks the Institute for Advanced Study for their hospitality and abundance of baked goods.
S.G. acknowledges support from the Fulbright U.S. Student Program.  
L.H. acknowledges support by the DOE DE-SC0011941 and a Simons Fellowship in Theoretical Physics.
J.C.H. acknowledges support from NSF grant AST-2108536 and NASA grant 21-ATP21-0129.  The Flatiron Institute is supported by the Simons Foundation.

\clearpage

\appendix

\section{Verification of the squeezed bispectrum in perturbation theory}\label{appen: sq-bk-loop}

In this Appendix, we verify the form of the squeezed bispectrum (Eq.~\ref{eq: sq-Bk-fNL}) using Eulerian perturbation theory, assuming $q \ll k$, with $k$ in the quasi-linear regime. We work in the infinite volume limit for clarity. Since we are interested in the leading $1/q^2$ term, which is not generated by the non-linear gravitational evolution alone, we focus only on the contribution to the bispectrum coming from primordial non-Gaussianity.

\subsection{Tree-level}

In the presence of local primordial non-Gaussianity, a non-trivial bispectrum is generated in addition to that from gravitational evolution. In particular, neglecting gravitational corrections, it is given by~\citep[e.g.,][]{Planck:2019kim},
\begin{align}
    \begin{split}  
        B_{111}(\bm q,\bm k) ={}& \big\langle\delta^{(1)}_{\bm q} \delta^{(1)}_{\bm k}\delta^{(1)}_{-\bm q - \bm k} \big\rangle^\prime \\
	    ={}& 2f_{\rm NL}\frac{\alpha(|\bm q + \bm k|)}{\alpha(q)\alpha(k)}P_L(q)P_L(k) \\
	    &+\text{2 perms.} \,,
    \end{split}
\end{align}
where $P_L$ is the linear power spectrum, $\delta^{(n)}$ the $n$-th order density field~\cite{Bernardeau:2001qr}, and by ${\langle\,\cdots\rangle}^\prime$ we indicate the ensemble-averaged correlator with overall momentum conserving $\delta$-function removed. In the squeezed limit, recalling that $\alpha(q)\sim q^2$ (Eq.~\ref{eq: alpha-q-def}), this takes the form
\begin{align}
    \begin{split} \label{eq: B111-sq}
	    B_{111}(\bm q,\bm k) ={}& \frac{4f_{\rm NL}}{\alpha(q)}P_L(q)P_L(k) + \mathcal{O}\big((q/k)^{-1}\big) \,.
    \end{split}
\end{align}
This matches Eq.~\eqref{eq: sq-Bk-tmp1} with $\partial P(k)/\partial \log\sigma_8^2\to P(k)$, which holds true at tree level.

\subsection{One-loop}

When including non-linear gravitational evolution, additional contributions are generated in the bispectrum. In particular, the one-loop contributions sourced by primordial non-Gaussianity are
\begin{align}
    \begin{split}
	    B_\text{1-loop}(\bm q,\bm k) ={}& \big\langle\delta^{(3)}_{\bm q}\delta^{(1)}_{\bm k}\delta^{(1)}_{-\bm q - \bm k} \big\rangle^\prime + \text{2 perms.} \\
	    &+\big\langle\delta^{(2)}_{\bm q}\delta^{(2)}_{\bm k} \delta^{(1)}_{-\bm q-\bm k}\big\rangle^\prime + \text{2 perms.}\\
	    & + \big\langle\delta^{(\rm ct)}_{\bm q}\delta^{(1)}_{\bm k}\delta^{(1)}_{-\bm q-\bm k} \big\rangle^\prime + \text{2 perms.} \,,
    \end{split}
\end{align}
where the final term is an ultraviolet counterterm required by renormalization, with $\delta^{(\rm ct)}_{\bm q} \equiv -c_s^2 q^2\delta^{(1)}_{\bm q}$, with $c_s$ a constant corresponding to the speed of sound in the hydrodynamical description of matter evolution. Each of these terms involve five linear density fields, which, after Wick contractions, involve the linear power spectrum, $\big\langle{\delta^{(1)}\delta^{(1)}}\big\rangle$, and the primordial bispectrum, $\big\langle{\delta^{(1)}\delta^{(1)}\delta^{(1)}}\big\rangle$ (except for the counterterm, which involves a primordial bispectrum and a factor of $q^2$). The squeezed bispectrum is dominated by terms involving primordial bispectra of the form $B_{111}(\bm q,\bm k)\sim P_L(q)P_L(k)/\alpha(q)$, for some $\bm k$ satisfying $q\ll k$; any other terms are suppressed by powers of $q^2$. There are seven such contributions:
\begin{widetext}
    \begin{align}
        \begin{split}
	        B_\text{1-loop}(\vq,\vk) \supset{}& 4\int_{\vp}F_2(\vp,\vk-\vp)F_2(\vp+\vq,\vk-\vp)  B_{111}(\vq,\vp) P_L(|\vk-\vp|) + 3B_{111}(\vq,\vk)\int_{\vp}F_3(\vk,\vp,-\vp)P_L(p) \\
	        &+3B_{111}(\vq,\vk)\int_{\vp}F_3(-\vk-\vq,\vp,-\vp)P_L(p) + 3P_L(|\vq+\vk|)\int_{\vp}F_3(\vk+\vq,\vp,-\vp-\vq)B_{111}(\vq,\vp) \\
	        & + 3P_L(k)\int_{\vp}F_3(-\vk,\vp,-\vp-\vq)B_{111}(\vq,\vp) - c_s^2 k^2  B_{111}(\vq,\vk) - c_s^2 |\vq + \vk|^2 B_{111}(\vq,-\vq-\vk) \,,
        \end{split}
    \end{align}
\end{widetext}
utilizing momentum conservation, and denoting $\int_{\vp}\equiv \int d^3\vp/(2\pi)^3$. The $F_n$ functions are the standard perturbative kernels~\cite{Bernardeau:2001qr}. Terms three, five, and seven are simply terms two, four, and six with the hard momenta switched.

\begin{figure*}
\includegraphics[width=\textwidth]{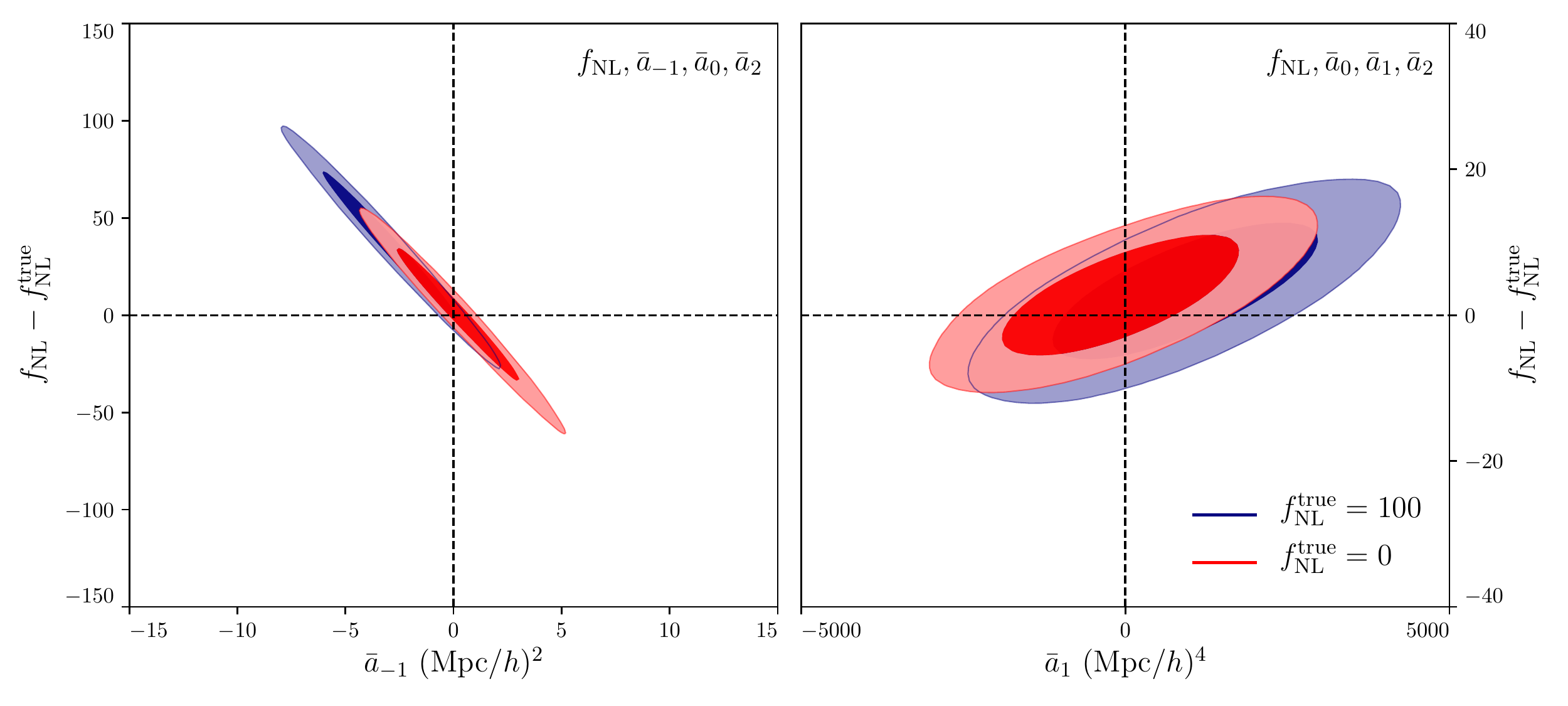}
\caption{Marginalized posteriors when including either $\bar{a}_{-1}$ ({\bf left}) or $\bar{a}_1$ ({\bf right}) as free parameters in the model. The odd coefficients are consistent with zero as is expected from our angular averaging procedure. There is significant degeneracy between $\overline{a}_{-1}$ and $f_{\rm NL}$ due to the difficulty in distinguishing between a $1/q$ and $1/q^2$ pole in our measurements.} \label{fig:odd_parameter_posteriors}
\end{figure*}

Taking the soft limit, $q\ll k,p$, and inserting Eq.~\eqref{eq: B111-sq}, these simplify to
\begin{widetext}
    \begin{align}
        \begin{split}
	      B_\text{1-loop}(\vq,\vk) ={}& 4 f_{\rm NL} \frac{P_L(q)}{\alpha(q)} \bigg[ 4  \int_{\vp}\left|F_2(\vp,\vk-\vp)\right|^2  P_L(p)P_L(|\vk-\vp|) + 12P_L(k) \int_{\vp}F_3(\vk,\vp,-\vp)P_L(p) \\
	        & - 2 c_s^2 k^2 P_L(k) \bigg] + \mathcal{O}\big((q/k)^{-1}\big) \,. 
        \end{split}
    \end{align}
\end{widetext}

We recognize the term in parentheses as twice the one-loop power spectrum, $P_\text{1-loop}(k)$~\cite{Baumann:2010tm,Carrasco:2012cv}. In combination with Eq.~\eqref{eq: B111-sq}, we find the full $f_{\rm NL}$ contribution to the squeezed bispectrum up to one-loop order:
\begin{align}\begin{split}
	B(\vq,\vk) ={}& \frac{4f_{\rm NL}}{\alpha(q)}P_L(q)\big[P_L(k) \\
	&+ 2P_\text{1-loop}(k)+\dots \big] + \mathcal{O}\big( (q/k)^{-1} \big) \,,
\end{split}\end{align}
which agrees with Eq.~\eqref{eq: sq-Bk-fNL}, noting that $P_L\propto \sigma_8^2$ and $P_\text{1-loop} \propto \sigma_8^4$.
 It is worth stressing that there is a factor of $2$ multiplying the one-loop contribution to the power spectrum. Thus what fits inside the square brackets is {\it not} simply the non-linear power spectrum.

\section{Validation of the parametrization} \label{app:validation}

In this Appendix, we perform a series of tests validating our fiducial analysis choice in which we vary $f_{\rm NL},\,\bar{a}_0,$ and $\bar{a}_2.$ In particular, we show that the odd coefficients in the power series expansion of the squeezed bispectrum are consistent with zero, that including the quadratic coefficient $\bar{a}_2$ is necessary to model the squeezed bispectrum at the scale cuts used in this work, and that coefficients higher than quadratic order can be neglected in our model. Unless otherwise stated, all results in this section use the mean measurements at redshift $z=0$ from 12 realizations with covariance scaled to match the volume of 12 realizations assuming a maximum soft momentum $q_{\rm max}=0.06~h/{\rm Mpc}$ and hard momenta between $0.2$ and $0.6~h/{\rm Mpc}$.

\begin{figure}[!b]
\includegraphics[width=\linewidth]{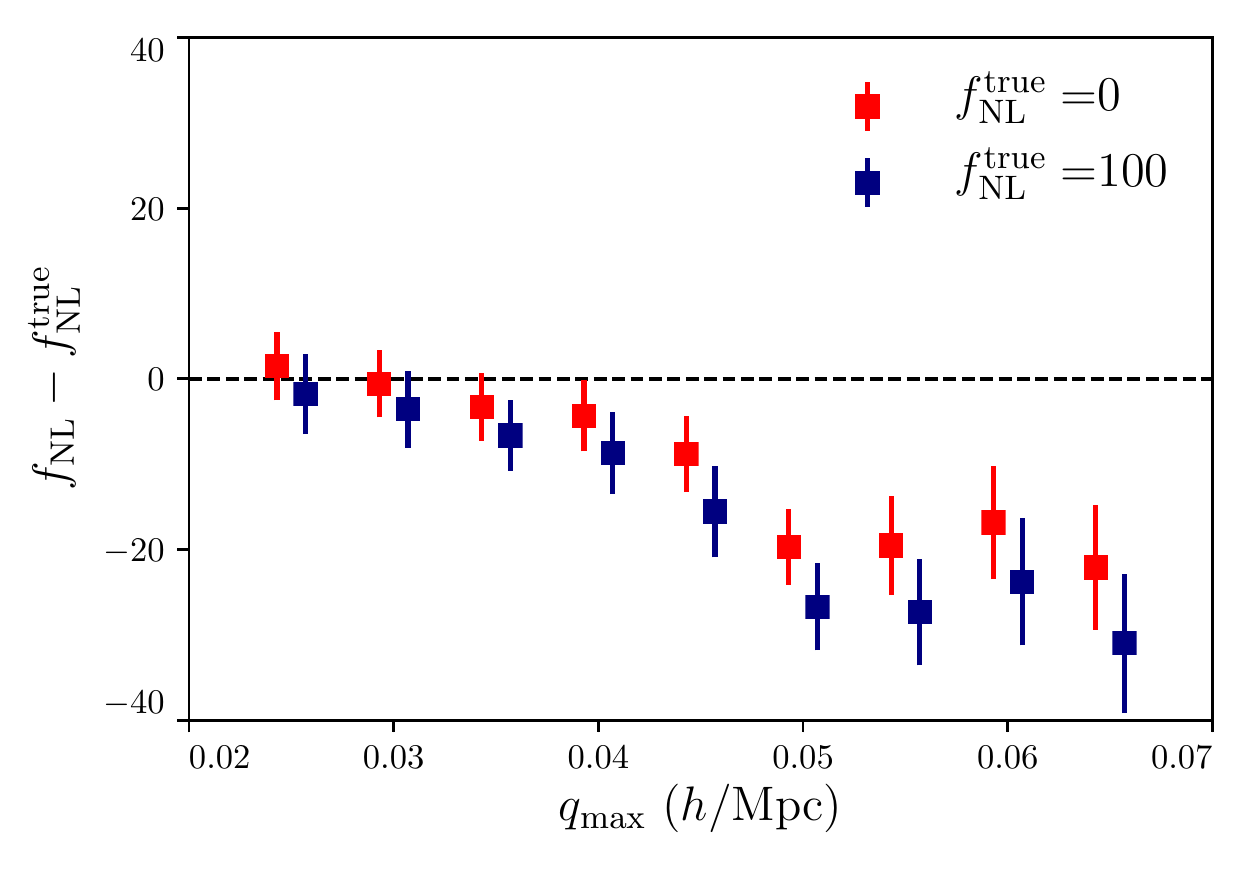}
\caption{Constraints on $f_{\rm NL}$ for varying soft momenta scale cuts assuming $\bar{a}_2=0$ and averaging over hard momenta between $0.2$ and $0.6~h/{\rm Mpc}$. Results are shown for $z=0$ for the simulations with Gaussian initial conditions (\textbf{red}), as well as those with primordial non-Gaussianity (\textbf{blue}). The horizontal offset between the two simulations is purely for visual purposes. We are unable to recover the true value of $f_{\rm NL}$ if we include soft momenta bins larger than $0.035~h/{\rm Mpc}$; therefore, we include $\bar{a}_2$ as a free parameter in our fiducial analysis.} \label{fig:qmax_dependence_2param}
\end{figure}

To show that $\bar{a}_{-1}$ and $\bar{a}_1$ can be set to zero (matching parity considerations) we rerun our analysis by including either $\bar{a}_{-1}$ or $\bar{a}_{1}$ as a free parameters.\footnote{We have also varied $\bar{a}_{-1}$ and $\bar{a}_1$ simultaneously and our results remain unchanged; however, the constraints degrade significantly when letting all parameters vary, thus we focus on the results for which we vary only one coefficient at a time.} In Figure~\ref{fig:odd_parameter_posteriors} we plot the 2D marginalized posteriors of $f_{\rm NL}-f_{\rm NL}^{\rm true}$ and the odd order coefficients. We include results with Gaussian initial conditions, as well as those assuming $f_{\rm NL}=100.$ The posteriors on $\bar{a}_{-1}$ and $\bar{a}_1$ are consistent with zero which suggests that our angular averaging is working as expected. Furthermore, including the odd coefficients as free parameters in our model does not bias the constraints on $f_{\rm NL}$. We note, however, that the error on $f_{\rm NL}$ increases significantly when we vary $\bar{a}_{-1}$, suggesting that our measurements are unable to fully differentiate between a $1/q$ and $1/q^2$ pole in the squeezed bispectrum. On account of these tests, we conclude that we can set the odd parameters in our model to zero.

\begin{figure}[!t]
\includegraphics[width=\linewidth]{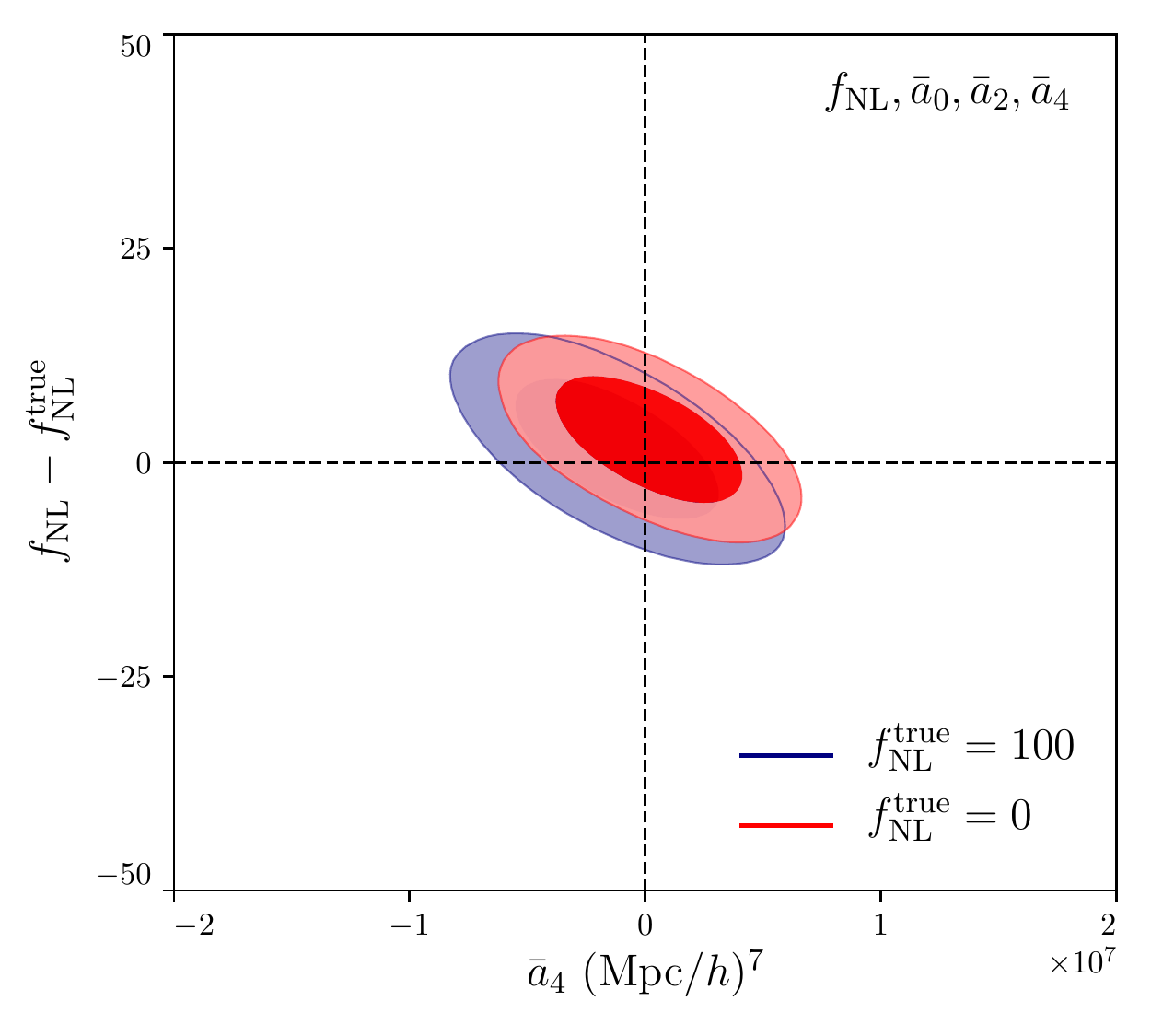}
\caption{Constraints on $f_{\rm NL}$ and $\bar{a}_4$ when including $\bar{a}_4$ as a free parameter in our model. $\bar{a}_4$ is consistent with zero, suggesting that it is sufficient to only consider terms up to $q^2$ in our fiducial analysis.} \label{fig:a4_posterior}
\end{figure}

In Figure~\ref{fig:qmax_dependence_2param} we instead plot the 68\% confidence limits on $f_{\rm NL}$ as a function of the maximum soft momentum, $q_{\rm max}$, assuming $\bar{a}_2=0$. As one can see, for $q_{\rm max} \gtrsim 0.035$~$h$/Mpc, this model gives biased constraints on $f_{\rm NL}$. Indeed, at these scales the $\mathcal{O}\big((q/k)^2\big)$ correction to the squeezed bispectrum becomes comparable to the measurement error. Hence, we need to include higher-order terms if we wish to include these modes in our analysis without increasing $k_{\rm min}$. Including $\bar{a}_2$ as a free parameter provides unbiased constraints on $f_{\rm NL}$ up to scales of at least $q_{\rm max}\simeq 0.065~h/{\rm Mpc}$ as shown in Figure~\ref{fig:kmin_dependence}. In practice one could either include $\bar{a}_2$ as a free parameter, or impose a stricter soft momentum scale cut and fix $\bar{a}_2=0$. In our fiducial analysis we choose to vary $\bar{a}_2$, since this leads to an error on $f_{\rm NL}$ that is approximately $20\%$ smaller than what we find when fixing $\bar{a}_2=0$ and decreasing $q_{\rm max}$. Alternatively, one could extend the validity of the two-parameter model to include more soft momenta bins by increasing $k_{\rm min};$ however, since the maximum accessible value of $k$ is often limited both in observations and simulations (e.g. by shot-noise or resolution) we do not utilize this method.

Finally, we rerun our analysis including terms up to quartic order in $q$ by varying $f_{\rm NL}, \bar{a}_0, \bar{a}_2,$ and $\bar{a}_4$. In Figure~\ref{fig:a4_posterior} we show the 2D marginalized posteriors of $f_{\rm NL}-f_{\rm NL}^{\rm true}$ and $\bar{a}_4$ . We find that $\bar{a}_4$ is completely consistent with zero and that we continue to recover unbiased constraints on $f_{\rm NL}$. Including $\bar{a}_4$ as a free parameter increases the error on $f_{\rm NL}$ by 50\%, thus we fix $\bar{a}_4$ and all higher-order coefficients to zero in our fiducial analysis.

\section{Likelihood and Fisher matrix}\label{appen: likelihood_fisher}

In this Appendix we derive the likelihood and Fisher matrix used in our analysis. We first start on general grounds, and then introduce simplifying approximations to ease the computational burden.

Consider the angular averaged bispectrum, $B_{iJ}$, binned in both soft and hard momenta bins. Here $i,j,\dots$ label the soft modes and $I,J,\dots$ the hard ones. Consistency relations imply that, in the squeezed limit, the angle-averaged bispectrum is a stochastic variable that fluctuates around its true value. For sufficiently thin bins, the latter is given by $\bar B_{iI} = s_{iI} \bar P_i$, where $\bar P_i$ is the true (soft) power spectrum, and $s_{iI}$ is defined as
\begin{align}
    s_{iI} \equiv f_{\rm NL} \, p_{iI} + \sum_{n \geq 0} a_{n,I} \, q_i^n \,,
\end{align}
with
\begin{align}
    p_{iI} \equiv \frac{6 \Omega_{m,0} H_0^2}{D_{\rm md}(z) T_i q_i^2} \frac{\partial \bar P_I}{\partial \log \sigma_8^2} \,.
\end{align}
Similarly, the power spectrum $P_i$ is a stochastic variable fluctuating around $\bar P_i$.

Assuming that both $B_{iJ}$ and $P_i$ are Gaussian distributed, we interpret their probability distribution as the likelihood for the unknown parameters $(f_{\rm NL}, \bm{a}_{n}, \bar P_i)$, which reads
\begin{widetext} 
    \begin{align} \label{eq:completeL}
        \begin{split}
            \mathcal{L}(f_{\rm NL}, \bm{a}_{n}, \bar P_i) \propto{}& \exp \bigg\{ - \frac{1}{2} \sum_{i} \Big[ (\bm{B}_{i} - \bm{s}_{i} \bar P_i) \cdot \bm{\Psi}^{BB}_{i} \cdot (\bm{B}_{i} - \bm{s}_{i} \bar P_i) + 2 (\bm{B}_i - \bm{s}_i \bar P_i) \cdot \bm{\Psi}^{BP}_{i} (P_i - \bar P_i) \\
             & + \Psi_{i}^{PP} \big(P_i - \bar P_i\big)^2 \Big] \bigg\} \,,
        \end{split}
    \end{align}
\end{widetext}
where, contrary to the main text, we explicitly let the vectors run over the hard mode indices, $I,J$. We have also used the fact that the covariance matrices are approximately diagonal in the soft mode, as verified in our simulations. Using the standard identities for the inverse of a block matrix, the $\Psi$'s can be related to the covariance matrices by,
\begin{subequations}
    \begin{align}
        \bm{\Psi}_i^{BB} ={}& \left( \bm{\mathcal{C}}_i^{BB} - \frac{\bm{\mathcal C}_i^{BP}\otimes \bm{\mathcal C}_i^{BP}}{\mathcal{C}_i^{PP}} \right)^{-1} \,, \\
        \bm{\Psi}_i^{BP} ={}& - \frac{\big(\bm{\mathcal C}_i^{BB}\big)^{-1}\cdot \bm{\mathcal C}_i^{BP}}{\mathcal{C}_i^{PP} - \bm{\mathcal C}_i^{BP} \cdot \big(\bm{\mathcal C}_i^{BB}\big)^{-1} \cdot \bm{\mathcal C}_i^{BP}} \,, \\
        \Psi_i^{PP} ={}& \frac{1}{\mathcal{C}_i^{PP} - \bm{\mathcal C}_i^{BP} \cdot \big(\bm{\mathcal C}_i^{BB}\big)^{-1} \cdot \bm{\mathcal C}_i^{BP}} \,,
    \end{align}
\end{subequations}
where ${\mathcal{C}}_{iIJ}^{BB}\equiv \langle \delta {B}_{iI}\delta {B}_{iJ}\rangle$, ${\mathcal{C}}_{iI}^{BP}\equiv \langle \delta {B}_{iI}\delta {P}_{i}\rangle$, ${\mathcal{C}}_{i}^{PP}\equiv \langle \delta {P}_{i}\delta {P}_{i}\rangle$, and $\langle \, \cdots \rangle$ denotes an ensemble average, which will be estimated from simulations. In particular, in the space of the hard momenta bins, $\bm{\Psi}_i^{BB}$ is a matrix, $\bm{\Psi}_i^{BP}$ a vector, and $\Psi_i^{PP}$ a scalar.

\subsection{The likelihood for the analysis}
\label{app:likelihood}

As described in Section~\ref{subsec:measurement_procedure}, in our main analysis we average over all hard modes collected in a single $k$-bin. The corresponding likelihood should then be obtained from Eq.~\eqref{eq:completeL} when the vector index is trivial, e.g., $\bm{a}_n \to \bar{a}_n$. Since we do not fit for the true value of the power spectrum, we also marginalize over $\bar P_i$, obtaining,
\begin{align} \label{eq:Lmarg}
    \mathcal{L}(f_{\rm NL},\bar{a}_n) \propto{}& \prod_i \frac{1}{\sqrt{\mathcal{C}_i}}\exp \left[ -\frac{1}{2} \frac{{(B_i - s_i P_i)}^2}{\mathcal{C}_i} \right] \,,
\end{align}
with $\mathcal{C}_i \equiv \mathcal{C}^{BB}_i - 2 s_i \mathcal{C}^{BP}_i + s_i^2 \mathcal{C}^{PP}_i$. Note that the likelihood above is simply a Gaussian one, but with a parameter dependent covariance.

In the analysis described in Section~\ref{Sec: likelihood} we start from the likelihood above, but introduce two modifications for practical purposes. First, since we are estimating the covariance matrix from simulations, we promote it to a non-diagonal one. We have checked that, even assuming a diagonal covariance in our analysis does not change the results appreciably. Second, in principle the only independent stochastic quantity in the theory model is the soft power spectrum, $P_i$. However, when accounting for finite size bins for which  $\big\langle q^n\delta_{\bm q}\delta_{-\bm q} \big\rangle\neq q^n\big\langle \delta_{\bm q}\delta_{-\bm q} \big\rangle$, the different terms in $s_i$ will weight $P_i$ by different powers of $q_i$, with effects that are hard to predict. We therefore find it more convenient to treat the $M_{in}$---defined in Eq.~\eqref{eq:Mhat}---as different stochastic variables. In summary, we employ the likelihood Eq.~\eqref{eq:Lmarg}, but promoting the covariance matrix to
\begin{align}
    \mathcal{C}_i \delta_{ij}^{\rm K} \to \Big\langle \Big(B_i - \sum_n \bar{a}_n M_{in}\Big)\Big(B_j - \sum_n \bar{a}_n M_{jn}\Big) \Big\rangle \,,
\end{align}
which, for a finite number of realizations, results in Eq.~\eqref{eq:likelihood}.

\subsection{The Fisher matrix}
\label{appen: fisher_matrix}

To determine the Fisher matrix we work with the original likelihood, Eq.~\eqref{eq:completeL}. To reduce the computational burden and reliably estimate the Fisher matrix from 40 mocks, we assume $\bm{a}_2=\bm{0}$ and hence the model parameters are $(f_{\rm NL}, \bm{a}_0, \bar P_i).$ The Fisher matrix is then given by the Hessian of $-\log \mathcal{L}$, and can be written in a block form as,
\begin{align}
    \bm{\mathcal{F}} = \begin{pmatrix}
                    \bm{F}_1 & \bm{F}_2 \\
                    \bm{F}_2^T & \bm{F}_3
                  \end{pmatrix} \,.
\end{align}
If we have $N_k$ hard mode bins, then $\bm{F}_1$ is a $(N_k + 1)\times(N_k+1)$ matrix with entries,
\begin{align}
    \bm{F}_1 \equiv \sum_i \bar P_i^2 \begin{pmatrix}
                                           \bm{p}_i\cdot\bm{\Psi}_i^{BB}\cdot\bm{p}_i &   \bm{p}_i\cdot\bm{\Psi}_i^{BB} \\
                                       \bm{\Psi}_i^{BB}\cdot\bm{p}_i & \bm{\Psi}_i^{BB} 
                                      \end{pmatrix} \,.
\end{align} 
If we have $N_q$ soft bins, $\bm{F}_2$ is a $(N_k+1)\times N_q$ matrix with entries,
\begin{align}
    (\bm{F}_2)_i \equiv \bar{P}_i \begin{pmatrix}
                               \bm{p}_i\cdot\bm{\Psi}_i^{BB}\cdot\bm{s}_i + \bm{p}_i\cdot\bm{\Psi}_i^{BP} \\
                               \bm{\Psi}_i^{BB}\cdot\bm{s}_i + \bm{\Psi}_i^{BP} 
                              \end{pmatrix} \,.
\end{align}
Finally, $\bm{F}_3$ is a diagonal $N_q\times N_q$ matrix,
\begin{align}
    (\bm{F}_3)_{ij} = \big( \bm{s}_i \cdot \bm{\Psi}_i^{BB} \cdot \bm{s}_i + 2 \bm{s}_i \cdot \bm{\Psi}_i^{BP} + \Psi_i^{PP} \big) \delta_{ij}^{\rm K} \,.
\end{align}
By the Cram\'er--Rao theorem, the variance on a measurement of $f_{\rm NL}$ will then satisfy the bound,
\begin{align}
    \sigma_{f_{\rm NL}}^2 \geq \big( \bm{\mathcal{F}}^{-1} \big)_{11} \,,
\end{align}
which is what we use in our analysis of Section~\ref{subsec: forecasts}. For the forecasts in Section~\ref{subsec: forecasts} we assume $f_{\rm NL}=0$ and that $\bar P_i$ is equal the mean power spectrum as measured from the 40 realizations with Gaussian initial conditions. To determine a set of fiducial values for $\bm{a}_0$ we must account for its $k$-dependence. To do this, we use the consistency relation from~\cite{Nishimichi:2014jna},
\begin{align}\label{eq:a0_CR}
    a_0(k)=\bigg[1+\frac{13}{21}\frac{\partial}{\partial \ln D_+}-\frac{1}{3}\frac{\partial}{\partial \ln k}\bigg]P(k),
\end{align}
where $D_+$ is the linear growing mode. In particular, we estimate $\bm{a}_0$ by computing Eq.~\eqref{eq:a0_CR} using \textsc{halofit}. This prediction agrees within 5\% of our measured value of $\bar{a}_0$ over all redshifts and scale cuts analyzed in this work, and is therefore sufficient for the forecasts.

\section{Optimal weighting schemes}\label{appen: weights}
In this Appendix, we consider possible optimal weighting schemes, $W(\vq, \vk)$, for the compressed bispectrum, $B(\vq) \equiv \int_{\vk}W(\vq, \vk)B(\vq, \vk)$. Note that in our main analysis we always fix $W(\vq,\vk)=1$ as we find that the weights derived in this section do not noticeably improve our constraints on $f_{\rm NL}$; nevertheless, we present our main findings below since they could be relevant for future works measuring $f_{\rm NL}$ from consistency relations. To derive the weights, we will compute the maximum-likelihood estimator for the $\bar{a}_n$ coefficients appearing in the bispectrum model of Eq.~\eqref{eq: theory_model} for a given $q$-bin using a simplified treatment of our likelihood and covariance.

For simplicity, we assume a Gaussian likelihood for the bispectrum and work in the continuous limit of $\vq$ and $\vk$.  The likelihood of the model parameters $\bar{a}_n$ is then, up to a constant term,
\begin{align}\label{eq: full-lik}
	\nonumber\log \mathcal{L}(\bar{a}_n) \propto -\int_{\vq\,\vq'\,\vk\,\vk'} \bigg(&\delta B(\bar{a}_n; \vq,\vk)\times\mathcal{C}^{-1}_{B}(\vq,\vk;\vq',\vk')\\ 
	&\times\delta B(\bar{a}_n;\vq',\vk')\bigg)\,,
\end{align}
where $\delta B(\bar{a}_n; \vq,\vk)\equiv B(\vq,\vk)-\sum_n\bar{a}_n\hat{M}_n(\vq)$ with $\hat{M}_n(\vq)$ defined via Eq.~\eqref{eq:Mhat}, $\mathcal{C}_B$ is the squeezed bispectrum covariance, and the integral is carried out over all soft and hard modes allowed for a given choice of scale cuts. In the squeezed limit, the bispectrum covariance is approximately given by \cite{Biagetti:2021tua} 
\begin{align}\label{eq: cov-B-approx}
	\mathcal{C}_B(\vq,\vk;\vk',\vq') &\simeq (2\pi)^3\delta_{\rm D}(\vq-\vq')\\\nonumber
	&\quad \times\left[P(\vq)\,\mathcal{C}_P(\vk,\vk')+2B(\vq',\vk')B(\vq,\vk)\right]\,,
\end{align}
where $\mathcal{C}_P$ is the (non-linear) power spectrum covariance.

Under null assumptions, i.e. $\bar{a}_n=0$ for all $n$, the bispectrum vanishes and only the first term survives, and the likelihood for a single $\vq$ mode becomes
\begin{align}\label{eq: lik-q}
	\log \mathcal{L}(\bar{a}_n, \vq) \propto -\frac{1}{P(\vq)}\int_{\vk\,\vk'}\bigg(&\delta B(\bar{a}_n; \vq,\vk)\times\mathcal{C}^{-1}_{P}(\vk,\vk') \nonumber \\
	&\times\delta B(\bar{a}_n; \vq',\vk')\bigg) \,,
\end{align}
up to a constant term. Given Eq.~\eqref{eq: lik-q}, we may derive an optimal estimator for $\bar{a}_n$ (for a single $\vq$-mode) by maximizing $\log \mathcal{L}(\bar{a}_n, \vq)$ with respect to $\bar{a}_n$. This yields
\begin{align}\begin{split}
	\widehat{\bar{a}_n} = \sum_m\mathcal{K}_{nm}^{-1}\int_{\vk\,\vk'}B(\vq,\vk)\,\mathcal{C}_P^{-1}(\vk,\vk')\,\hat{M}_m(\vq) \,,
\end{split}\end{align}
with the coupling matrix 
\begin{align}\begin{split}
	\mathcal{K}_{mn} =\int_{\vk\,\vk'}\hat{M}_m(\vq)\,\mathcal{C}_P^{-1}(\vk,\vk')\,\hat{M}_n(\vq) \,.
\end{split}\end{align}
The optimal weight is thus $W_n(\vq,\vk) \propto \int_{\vk'}\mathcal{C}_P^{-1}(\vk,\vk')\hat{M}_n(\vq)$, with one weight per choice of $n$. 

It is useful to consider two simplifying limits. First, if the covariance is dominated by its Gaussian contribution, $V\mathcal{C}_P(\vk,\vk') = 2(2\pi)^3\delta_{\rm D}(\vk-\vk')P^2(k)$, yielding
\begin{align}\begin{split}
	\widehat{a_n} &\to \sum_m\mathcal{K}_{nm}^{-1}\int_{\vk}\frac{B(\vq,\vk)\hat{M}_m(\vq)}{P^2(k)} \,, \\
	\mathcal{K}_{mn} &\to\int_{\vk}\frac{\hat{M}_m(\vq)\hat{M}_n(\vq)}{P^2(k)} \,;
\end{split}\end{align}
this allows us to remove the $\vk'$ integral. Second, if the coupling between $\bar{a}_n$ coefficients is weak (i.e. $\mathcal{K}_{mn}$ is close to diagonal), we may apply the optimal weighting only for the $m=-2$ coefficient. Inserting the relevant template (Eq.~\ref{eq: am2-fNL}), we find the quasi-optimal weighting $W(\vq,\vk) = \partial \log P(k)/\partial\log\sigma_8^2 / P(k)$ (dropping $\vk$-independent factors), and thus the compressed bispectrum
\begin{align}\begin{split}
	B(\vq) \equiv \int_{\vk} \frac{B(\vq,\vk)}{P(k)} \frac{\partial \log P(k)}{\partial\log\sigma_8^2}\,.
\end{split}\end{align}
Strictly speaking, this weighting could be far from optimal in the high-$k$ limit, whereupon the power spectrum covariance is far from diagonal. In this case, invoking only the second assumption, one might use
\begin{align}\begin{split}
	B(\vq) \equiv \int_{\vk} B(\vq,\vk)\int_{\vk'}\mathcal{C}^{-1}_P(\vk,\vk')\frac{\partial P(k')}{\partial\log\sigma_8^2}\,.
\end{split}\end{align}
This may be of greater use in practice, and is straightforward to implement if the weighting is estimated from simulations, e.g., \textsc{Quijote}. Finally, one may wish to retain the bispectra appearing in Eq.~\eqref{eq: cov-B-approx}, assuming non-zero fiducial values of $\bar{a}_n$: this can be done similarly, though the covariance will become a function also of $q$.

\section{Full posterior distribution of fiducial analysis} \label{app:posterior}

\begin{figure*}
    \centering
    \includegraphics[width=\linewidth]{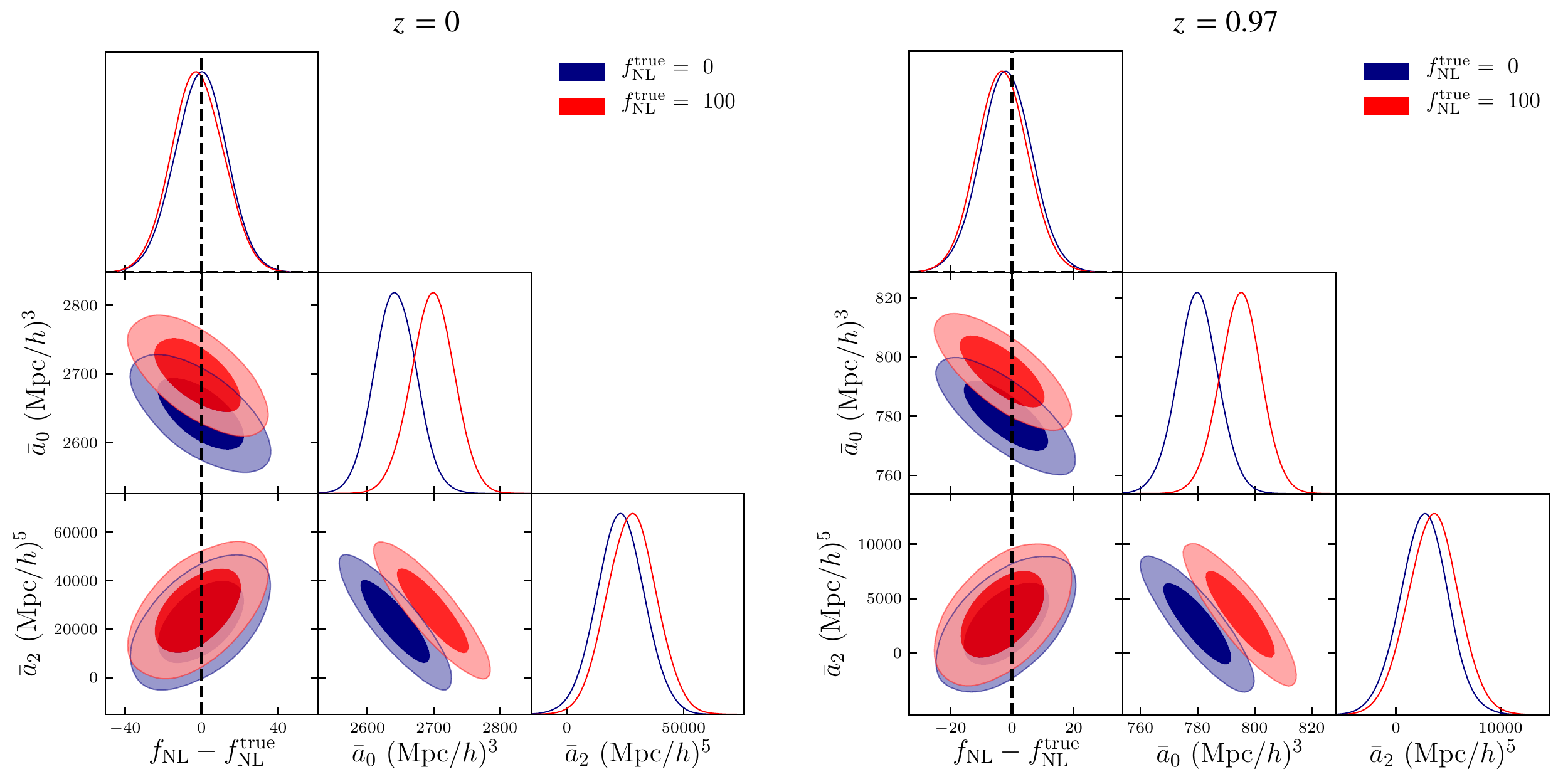}%
    \caption{Full marginalized posteriors of the fiducial results presented in Section~\ref{sec: results} at $z=0$ ({\bf left}) and $z=0.97$ ({\bf right}). We recover the true value of $f_{\rm NL}$ in all cases with considerable improvement at $z=0.97.$}\label{fig:full_posterior}
\end{figure*}

In Figure~\ref{fig:full_posterior} we show the full marginalized posteriors of our fiducial analysis at $z=0$ and $z=0.97$. As seen in the text, we recover the true value of $f_{\rm NL}$ for both redshifts and both sets of initial conditions. There are noticeable degeneracies between all parameters, with a particularly strong correlation between $\bar{a}_0$ and $\bar{a}_2.$ We also find that $\bar{a}_0$ is larger for simulations with $f_{\rm NL}=100$. This can be understood by noting that $\bar{a}_0 \sim P(k)$ and a positive value of $f_{\rm NL}$ enhances the non-linear power spectrum \cite{Taruya:2008pg}.

\bibliographystyle{apsrev4-1}
\bibliography{biblio}

\begin{thebibliography}{68}%
\makeatletter
\providecommand \@ifxundefined [1]{%
 \@ifx{#1\undefined}
}%
\providecommand \@ifnum [1]{%
 \ifnum #1\expandafter \@firstoftwo
 \else \expandafter \@secondoftwo
 \fi
}%
\providecommand \@ifx [1]{%
 \ifx #1\expandafter \@firstoftwo
 \else \expandafter \@secondoftwo
 \fi
}%
\providecommand \natexlab [1]{#1}%
\providecommand \enquote  [1]{``#1''}%
\providecommand \bibnamefont  [1]{#1}%
\providecommand \bibfnamefont [1]{#1}%
\providecommand \citenamefont [1]{#1}%
\providecommand \href@noop [0]{\@secondoftwo}%
\providecommand \href [0]{\begingroup \@sanitize@url \@href}%
\providecommand \@href[1]{\@@startlink{#1}\@@href}%
\providecommand \@@href[1]{\endgroup#1\@@endlink}%
\providecommand \@sanitize@url [0]{\catcode `\\12\catcode `\$12\catcode
  `\&12\catcode `\#12\catcode `\^12\catcode `\_12\catcode `\%12\relax}%
\providecommand \@@startlink[1]{}%
\providecommand \@@endlink[0]{}%
\providecommand \url  [0]{\begingroup\@sanitize@url \@url }%
\providecommand \@url [1]{\endgroup\@href {#1}{\urlprefix }}%
\providecommand \urlprefix  [0]{URL }%
\providecommand \Eprint [0]{\href }%
\providecommand \doibase [0]{http://dx.doi.org/}%
\providecommand \selectlanguage [0]{\@gobble}%
\providecommand \bibinfo  [0]{\@secondoftwo}%
\providecommand \bibfield  [0]{\@secondoftwo}%
\providecommand \translation [1]{[#1]}%
\providecommand \BibitemOpen [0]{}%
\providecommand \bibitemStop [0]{}%
\providecommand \bibitemNoStop [0]{.\EOS\space}%
\providecommand \EOS [0]{\spacefactor3000\relax}%
\providecommand \BibitemShut  [1]{\csname bibitem#1\endcsname}%
\let\auto@bib@innerbib\@empty
\bibitem [{\citenamefont {Maldacena}(2003)}]{Maldacena:2002vr}%
  \BibitemOpen
  \bibfield  {author} {\bibinfo {author} {\bibfnamefont {J.~M.}\ \bibnamefont
  {Maldacena}},\ }\href {\doibase 10.1088/1126-6708/2003/05/013} {\bibfield
  {journal} {\bibinfo  {journal} {JHEP}\ }\textbf {\bibinfo {volume} {05}},\
  \bibinfo {pages} {013} (\bibinfo {year} {2003})},\ \Eprint
  {http://arxiv.org/abs/astro-ph/0210603} {arXiv:astro-ph/0210603} \BibitemShut
  {NoStop}%
\bibitem [{\citenamefont {Creminelli}\ and\ \citenamefont
  {Zaldarriaga}(2004)}]{Creminelli:2004yq}%
  \BibitemOpen
  \bibfield  {author} {\bibinfo {author} {\bibfnamefont {P.}~\bibnamefont
  {Creminelli}}\ and\ \bibinfo {author} {\bibfnamefont {M.}~\bibnamefont
  {Zaldarriaga}},\ }\href {\doibase 10.1088/1475-7516/2004/10/006} {\bibfield
  {journal} {\bibinfo  {journal} {JCAP}\ }\textbf {\bibinfo {volume} {10}},\
  \bibinfo {pages} {006} (\bibinfo {year} {2004})},\ \Eprint
  {http://arxiv.org/abs/astro-ph/0407059} {arXiv:astro-ph/0407059} \BibitemShut
  {NoStop}%
\bibitem [{\citenamefont {Cheung}\ \emph {et~al.}(2008)\citenamefont {Cheung},
  \citenamefont {Fitzpatrick}, \citenamefont {Kaplan},\ and\ \citenamefont
  {Senatore}}]{Cheung:2007sv}%
  \BibitemOpen
  \bibfield  {author} {\bibinfo {author} {\bibfnamefont {C.}~\bibnamefont
  {Cheung}}, \bibinfo {author} {\bibfnamefont {A.~L.}\ \bibnamefont
  {Fitzpatrick}}, \bibinfo {author} {\bibfnamefont {J.}~\bibnamefont {Kaplan}},
  \ and\ \bibinfo {author} {\bibfnamefont {L.}~\bibnamefont {Senatore}},\
  }\href {\doibase 10.1088/1475-7516/2008/02/021} {\bibfield  {journal}
  {\bibinfo  {journal} {JCAP}\ }\textbf {\bibinfo {volume} {02}},\ \bibinfo
  {pages} {021} (\bibinfo {year} {2008})},\ \Eprint
  {http://arxiv.org/abs/0709.0295} {arXiv:0709.0295 [hep-th]} \BibitemShut
  {NoStop}%
\bibitem [{\citenamefont {Tanaka}\ and\ \citenamefont
  {Urakawa}(2011)}]{Tanaka:2011aj}%
  \BibitemOpen
  \bibfield  {author} {\bibinfo {author} {\bibfnamefont {T.}~\bibnamefont
  {Tanaka}}\ and\ \bibinfo {author} {\bibfnamefont {Y.}~\bibnamefont
  {Urakawa}},\ }\href {\doibase 10.1088/1475-7516/2011/05/014} {\bibfield
  {journal} {\bibinfo  {journal} {JCAP}\ }\textbf {\bibinfo {volume} {05}},\
  \bibinfo {pages} {014} (\bibinfo {year} {2011})},\ \Eprint
  {http://arxiv.org/abs/1103.1251} {arXiv:1103.1251 [astro-ph.CO]} \BibitemShut
  {NoStop}%
\bibitem [{\citenamefont {Creminelli}\ \emph {et~al.}(2012)\citenamefont
  {Creminelli}, \citenamefont {Nore\~na},\ and\ \citenamefont
  {Simonovi\'c}}]{Creminelli:2012ed}%
  \BibitemOpen
  \bibfield  {author} {\bibinfo {author} {\bibfnamefont {P.}~\bibnamefont
  {Creminelli}}, \bibinfo {author} {\bibfnamefont {J.}~\bibnamefont
  {Nore\~na}}, \ and\ \bibinfo {author} {\bibfnamefont {M.}~\bibnamefont
  {Simonovi\'c}},\ }\href {\doibase 10.1088/1475-7516/2012/07/052} {\bibfield
  {journal} {\bibinfo  {journal} {JCAP}\ }\textbf {\bibinfo {volume} {07}},\
  \bibinfo {pages} {052} (\bibinfo {year} {2012})},\ \Eprint
  {http://arxiv.org/abs/1203.4595} {arXiv:1203.4595 [hep-th]} \BibitemShut
  {NoStop}%
\bibitem [{\citenamefont {Hinterbichler}\ \emph {et~al.}(2012)\citenamefont
  {Hinterbichler}, \citenamefont {Hui},\ and\ \citenamefont
  {Khoury}}]{Hinterbichler:2012nm}%
  \BibitemOpen
  \bibfield  {author} {\bibinfo {author} {\bibfnamefont {K.}~\bibnamefont
  {Hinterbichler}}, \bibinfo {author} {\bibfnamefont {L.}~\bibnamefont {Hui}},
  \ and\ \bibinfo {author} {\bibfnamefont {J.}~\bibnamefont {Khoury}},\ }\href
  {\doibase 10.1088/1475-7516/2012/08/017} {\bibfield  {journal} {\bibinfo
  {journal} {JCAP}\ }\textbf {\bibinfo {volume} {08}},\ \bibinfo {pages} {017}
  (\bibinfo {year} {2012})},\ \Eprint {http://arxiv.org/abs/1203.6351}
  {arXiv:1203.6351 [hep-th]} \BibitemShut {NoStop}%
\bibitem [{\citenamefont {Assassi}\ \emph {et~al.}(2012)\citenamefont
  {Assassi}, \citenamefont {Baumann},\ and\ \citenamefont
  {Green}}]{Assassi:2012zq}%
  \BibitemOpen
  \bibfield  {author} {\bibinfo {author} {\bibfnamefont {V.}~\bibnamefont
  {Assassi}}, \bibinfo {author} {\bibfnamefont {D.}~\bibnamefont {Baumann}}, \
  and\ \bibinfo {author} {\bibfnamefont {D.}~\bibnamefont {Green}},\ }\href
  {\doibase 10.1088/1475-7516/2012/11/047} {\bibfield  {journal} {\bibinfo
  {journal} {JCAP}\ }\textbf {\bibinfo {volume} {11}},\ \bibinfo {pages} {047}
  (\bibinfo {year} {2012})},\ \Eprint {http://arxiv.org/abs/1204.4207}
  {arXiv:1204.4207 [hep-th]} \BibitemShut {NoStop}%
\bibitem [{\citenamefont {Kehagias}\ and\ \citenamefont
  {Riotto}(2012)}]{Kehagias:2012pd}%
  \BibitemOpen
  \bibfield  {author} {\bibinfo {author} {\bibfnamefont {A.}~\bibnamefont
  {Kehagias}}\ and\ \bibinfo {author} {\bibfnamefont {A.}~\bibnamefont
  {Riotto}},\ }\href {\doibase 10.1016/j.nuclphysb.2012.07.004} {\bibfield
  {journal} {\bibinfo  {journal} {Nucl. Phys. B}\ }\textbf {\bibinfo {volume}
  {864}},\ \bibinfo {pages} {492} (\bibinfo {year} {2012})},\ \Eprint
  {http://arxiv.org/abs/1205.1523} {arXiv:1205.1523 [hep-th]} \BibitemShut
  {NoStop}%
\bibitem [{\citenamefont {Pajer}\ \emph {et~al.}(2013)\citenamefont {Pajer},
  \citenamefont {Schmidt},\ and\ \citenamefont {Zaldarriaga}}]{Pajer:2013ana}%
  \BibitemOpen
  \bibfield  {author} {\bibinfo {author} {\bibfnamefont {E.}~\bibnamefont
  {Pajer}}, \bibinfo {author} {\bibfnamefont {F.}~\bibnamefont {Schmidt}}, \
  and\ \bibinfo {author} {\bibfnamefont {M.}~\bibnamefont {Zaldarriaga}},\
  }\href {\doibase 10.1103/PhysRevD.88.083502} {\bibfield  {journal} {\bibinfo
  {journal} {Phys. Rev. D}\ }\textbf {\bibinfo {volume} {88}},\ \bibinfo
  {pages} {083502} (\bibinfo {year} {2013})},\ \Eprint
  {http://arxiv.org/abs/1305.0824} {arXiv:1305.0824 [astro-ph.CO]} \BibitemShut
  {NoStop}%
\bibitem [{\citenamefont {Hinterbichler}\ \emph {et~al.}(2014)\citenamefont
  {Hinterbichler}, \citenamefont {Hui},\ and\ \citenamefont
  {Khoury}}]{Hinterbichler:2013dpa}%
  \BibitemOpen
  \bibfield  {author} {\bibinfo {author} {\bibfnamefont {K.}~\bibnamefont
  {Hinterbichler}}, \bibinfo {author} {\bibfnamefont {L.}~\bibnamefont {Hui}},
  \ and\ \bibinfo {author} {\bibfnamefont {J.}~\bibnamefont {Khoury}},\ }\href
  {\doibase 10.1088/1475-7516/2014/01/039} {\bibfield  {journal} {\bibinfo
  {journal} {JCAP}\ }\textbf {\bibinfo {volume} {01}},\ \bibinfo {pages} {039}
  (\bibinfo {year} {2014})},\ \Eprint {http://arxiv.org/abs/1304.5527}
  {arXiv:1304.5527 [hep-th]} \BibitemShut {NoStop}%
\bibitem [{\citenamefont {Goldberger}\ \emph {et~al.}(2013)\citenamefont
  {Goldberger}, \citenamefont {Hui},\ and\ \citenamefont
  {Nicolis}}]{Goldberger:2013rsa}%
  \BibitemOpen
  \bibfield  {author} {\bibinfo {author} {\bibfnamefont {W.~D.}\ \bibnamefont
  {Goldberger}}, \bibinfo {author} {\bibfnamefont {L.}~\bibnamefont {Hui}}, \
  and\ \bibinfo {author} {\bibfnamefont {A.}~\bibnamefont {Nicolis}},\ }\href
  {\doibase 10.1103/PhysRevD.87.103520} {\bibfield  {journal} {\bibinfo
  {journal} {Phys. Rev. D}\ }\textbf {\bibinfo {volume} {87}},\ \bibinfo
  {pages} {103520} (\bibinfo {year} {2013})},\ \Eprint
  {http://arxiv.org/abs/1303.1193} {arXiv:1303.1193 [hep-th]} \BibitemShut
  {NoStop}%
\bibitem [{\citenamefont {Baldauf}\ \emph {et~al.}(2015)\citenamefont
  {Baldauf}, \citenamefont {Mirbabayi}, \citenamefont {Simonovi\'c},\ and\
  \citenamefont {Zaldarriaga}}]{Baldauf:2015xfa}%
  \BibitemOpen
  \bibfield  {author} {\bibinfo {author} {\bibfnamefont {T.}~\bibnamefont
  {Baldauf}}, \bibinfo {author} {\bibfnamefont {M.}~\bibnamefont {Mirbabayi}},
  \bibinfo {author} {\bibfnamefont {M.}~\bibnamefont {Simonovi\'c}}, \ and\
  \bibinfo {author} {\bibfnamefont {M.}~\bibnamefont {Zaldarriaga}},\ }\href
  {\doibase 10.1103/PhysRevD.92.043514} {\bibfield  {journal} {\bibinfo
  {journal} {Phys. Rev. D}\ }\textbf {\bibinfo {volume} {92}},\ \bibinfo
  {pages} {043514} (\bibinfo {year} {2015})},\ \Eprint
  {http://arxiv.org/abs/1504.04366} {arXiv:1504.04366 [astro-ph.CO]}
  \BibitemShut {NoStop}%
\bibitem [{\citenamefont {Bravo}\ \emph {et~al.}(2018)\citenamefont {Bravo},
  \citenamefont {Mooij}, \citenamefont {Palma},\ and\ \citenamefont
  {Pradenas}}]{Bravo:2017gct}%
  \BibitemOpen
  \bibfield  {author} {\bibinfo {author} {\bibfnamefont {R.}~\bibnamefont
  {Bravo}}, \bibinfo {author} {\bibfnamefont {S.}~\bibnamefont {Mooij}},
  \bibinfo {author} {\bibfnamefont {G.~A.}\ \bibnamefont {Palma}}, \ and\
  \bibinfo {author} {\bibfnamefont {B.}~\bibnamefont {Pradenas}},\ }\href
  {\doibase 10.1088/1475-7516/2018/05/025} {\bibfield  {journal} {\bibinfo
  {journal} {JCAP}\ }\textbf {\bibinfo {volume} {05}},\ \bibinfo {pages} {025}
  (\bibinfo {year} {2018})},\ \Eprint {http://arxiv.org/abs/1711.05290}
  {arXiv:1711.05290 [astro-ph.CO]} \BibitemShut {NoStop}%
\bibitem [{\citenamefont {Hui}\ \emph {et~al.}(2019)\citenamefont {Hui},
  \citenamefont {Joyce},\ and\ \citenamefont {Wong}}]{Hui:2018cag}%
  \BibitemOpen
  \bibfield  {author} {\bibinfo {author} {\bibfnamefont {L.}~\bibnamefont
  {Hui}}, \bibinfo {author} {\bibfnamefont {A.}~\bibnamefont {Joyce}}, \ and\
  \bibinfo {author} {\bibfnamefont {S.~S.~C.}\ \bibnamefont {Wong}},\ }\href
  {\doibase 10.1088/1475-7516/2019/02/060} {\bibfield  {journal} {\bibinfo
  {journal} {JCAP}\ }\textbf {\bibinfo {volume} {02}},\ \bibinfo {pages} {060}
  (\bibinfo {year} {2019})},\ \Eprint {http://arxiv.org/abs/1811.05951}
  {arXiv:1811.05951 [hep-th]} \BibitemShut {NoStop}%
\bibitem [{\citenamefont {Kehagias}\ and\ \citenamefont
  {Riotto}(2013)}]{Kehagias:2013yd}%
  \BibitemOpen
  \bibfield  {author} {\bibinfo {author} {\bibfnamefont {A.}~\bibnamefont
  {Kehagias}}\ and\ \bibinfo {author} {\bibfnamefont {A.}~\bibnamefont
  {Riotto}},\ }\href {\doibase 10.1016/j.nuclphysb.2013.05.009} {\bibfield
  {journal} {\bibinfo  {journal} {Nucl. Phys. B}\ }\textbf {\bibinfo {volume}
  {873}},\ \bibinfo {pages} {514} (\bibinfo {year} {2013})},\ \Eprint
  {http://arxiv.org/abs/1302.0130} {arXiv:1302.0130 [astro-ph.CO]} \BibitemShut
  {NoStop}%
\bibitem [{\citenamefont {Peloso}\ and\ \citenamefont
  {Pietroni}(2013)}]{Peloso:2013zw}%
  \BibitemOpen
  \bibfield  {author} {\bibinfo {author} {\bibfnamefont {M.}~\bibnamefont
  {Peloso}}\ and\ \bibinfo {author} {\bibfnamefont {M.}~\bibnamefont
  {Pietroni}},\ }\href {\doibase 10.1088/1475-7516/2013/05/031} {\bibfield
  {journal} {\bibinfo  {journal} {JCAP}\ }\textbf {\bibinfo {volume} {05}},\
  \bibinfo {pages} {031} (\bibinfo {year} {2013})},\ \Eprint
  {http://arxiv.org/abs/1302.0223} {arXiv:1302.0223 [astro-ph.CO]} \BibitemShut
  {NoStop}%
\bibitem [{\citenamefont {Valageas}(2014)}]{Valageas:2013zda}%
  \BibitemOpen
  \bibfield  {author} {\bibinfo {author} {\bibfnamefont {P.}~\bibnamefont
  {Valageas}},\ }\href {\doibase 10.1103/PhysRevD.89.123522} {\bibfield
  {journal} {\bibinfo  {journal} {Phys. Rev. D}\ }\textbf {\bibinfo {volume}
  {89}},\ \bibinfo {pages} {123522} (\bibinfo {year} {2014})},\ \Eprint
  {http://arxiv.org/abs/1311.4286} {arXiv:1311.4286 [astro-ph.CO]} \BibitemShut
  {NoStop}%
\bibitem [{\citenamefont {Creminelli}\ \emph
  {et~al.}(2014{\natexlab{a}})\citenamefont {Creminelli}, \citenamefont
  {Gleyzes}, \citenamefont {Simonovi\'c},\ and\ \citenamefont
  {Vernizzi}}]{Creminelli:2013poa}%
  \BibitemOpen
  \bibfield  {author} {\bibinfo {author} {\bibfnamefont {P.}~\bibnamefont
  {Creminelli}}, \bibinfo {author} {\bibfnamefont {J.}~\bibnamefont {Gleyzes}},
  \bibinfo {author} {\bibfnamefont {M.}~\bibnamefont {Simonovi\'c}}, \ and\
  \bibinfo {author} {\bibfnamefont {F.}~\bibnamefont {Vernizzi}},\ }\href
  {\doibase 10.1088/1475-7516/2014/02/051} {\bibfield  {journal} {\bibinfo
  {journal} {JCAP}\ }\textbf {\bibinfo {volume} {02}},\ \bibinfo {pages} {051}
  (\bibinfo {year} {2014}{\natexlab{a}})},\ \Eprint
  {http://arxiv.org/abs/1311.0290} {arXiv:1311.0290 [astro-ph.CO]} \BibitemShut
  {NoStop}%
\bibitem [{\citenamefont {Valageas}\ \emph {et~al.}(2017)\citenamefont
  {Valageas}, \citenamefont {Taruya},\ and\ \citenamefont
  {Nishimichi}}]{Valageas:2016hhr}%
  \BibitemOpen
  \bibfield  {author} {\bibinfo {author} {\bibfnamefont {P.}~\bibnamefont
  {Valageas}}, \bibinfo {author} {\bibfnamefont {A.}~\bibnamefont {Taruya}}, \
  and\ \bibinfo {author} {\bibfnamefont {T.}~\bibnamefont {Nishimichi}},\
  }\href {\doibase 10.1103/PhysRevD.95.023504} {\bibfield  {journal} {\bibinfo
  {journal} {Phys. Rev. D}\ }\textbf {\bibinfo {volume} {95}},\ \bibinfo
  {pages} {023504} (\bibinfo {year} {2017})},\ \Eprint
  {http://arxiv.org/abs/1610.00993} {arXiv:1610.00993 [astro-ph.CO]}
  \BibitemShut {NoStop}%
\bibitem [{\citenamefont {Marinucci}\ \emph {et~al.}(2019)\citenamefont
  {Marinucci}, \citenamefont {Nishimichi},\ and\ \citenamefont
  {Pietroni}}]{Marinucci:2019wdb}%
  \BibitemOpen
  \bibfield  {author} {\bibinfo {author} {\bibfnamefont {M.}~\bibnamefont
  {Marinucci}}, \bibinfo {author} {\bibfnamefont {T.}~\bibnamefont
  {Nishimichi}}, \ and\ \bibinfo {author} {\bibfnamefont {M.}~\bibnamefont
  {Pietroni}},\ }\href {\doibase 10.1103/PhysRevD.100.123537} {\bibfield
  {journal} {\bibinfo  {journal} {Phys. Rev. D}\ }\textbf {\bibinfo {volume}
  {100}},\ \bibinfo {pages} {123537} (\bibinfo {year} {2019})},\ \Eprint
  {http://arxiv.org/abs/1907.09866} {arXiv:1907.09866 [astro-ph.CO]}
  \BibitemShut {NoStop}%
\bibitem [{\citenamefont {Marinucci}\ \emph {et~al.}(2020)\citenamefont
  {Marinucci}, \citenamefont {Nishimichi},\ and\ \citenamefont
  {Pietroni}}]{Marinucci:2020weg}%
  \BibitemOpen
  \bibfield  {author} {\bibinfo {author} {\bibfnamefont {M.}~\bibnamefont
  {Marinucci}}, \bibinfo {author} {\bibfnamefont {T.}~\bibnamefont
  {Nishimichi}}, \ and\ \bibinfo {author} {\bibfnamefont {M.}~\bibnamefont
  {Pietroni}},\ }\href {\doibase 10.1088/1475-7516/2020/07/054} {\bibfield
  {journal} {\bibinfo  {journal} {JCAP}\ }\textbf {\bibinfo {volume} {07}},\
  \bibinfo {pages} {054} (\bibinfo {year} {2020})},\ \Eprint
  {http://arxiv.org/abs/2005.09574} {arXiv:2005.09574 [astro-ph.CO]}
  \BibitemShut {NoStop}%
\bibitem [{\citenamefont {Creminelli}\ \emph {et~al.}(2013)\citenamefont
  {Creminelli}, \citenamefont {Nore\~na}, \citenamefont {Simonovi\'c},\ and\
  \citenamefont {Vernizzi}}]{Creminelli:2013mca}%
  \BibitemOpen
  \bibfield  {author} {\bibinfo {author} {\bibfnamefont {P.}~\bibnamefont
  {Creminelli}}, \bibinfo {author} {\bibfnamefont {J.}~\bibnamefont
  {Nore\~na}}, \bibinfo {author} {\bibfnamefont {M.}~\bibnamefont
  {Simonovi\'c}}, \ and\ \bibinfo {author} {\bibfnamefont {F.}~\bibnamefont
  {Vernizzi}},\ }\href {\doibase 10.1088/1475-7516/2013/12/025} {\bibfield
  {journal} {\bibinfo  {journal} {JCAP}\ }\textbf {\bibinfo {volume} {12}},\
  \bibinfo {pages} {025} (\bibinfo {year} {2013})},\ \Eprint
  {http://arxiv.org/abs/1309.3557} {arXiv:1309.3557 [astro-ph.CO]} \BibitemShut
  {NoStop}%
\bibitem [{\citenamefont {Esposito}\ \emph {et~al.}(2019)\citenamefont
  {Esposito}, \citenamefont {Hui},\ and\ \citenamefont
  {Scoccimarro}}]{Esposito:2019jkb}%
  \BibitemOpen
  \bibfield  {author} {\bibinfo {author} {\bibfnamefont {A.}~\bibnamefont
  {Esposito}}, \bibinfo {author} {\bibfnamefont {L.}~\bibnamefont {Hui}}, \
  and\ \bibinfo {author} {\bibfnamefont {R.}~\bibnamefont {Scoccimarro}},\
  }\href {\doibase 10.1103/PhysRevD.100.043536} {\bibfield  {journal} {\bibinfo
   {journal} {Phys. Rev. D}\ }\textbf {\bibinfo {volume} {100}},\ \bibinfo
  {pages} {043536} (\bibinfo {year} {2019})},\ \Eprint
  {http://arxiv.org/abs/1905.11423} {arXiv:1905.11423 [astro-ph.CO]}
  \BibitemShut {NoStop}%
\bibitem [{\citenamefont {Kehagias}\ \emph {et~al.}(2014)\citenamefont
  {Kehagias}, \citenamefont {Nore\~na}, \citenamefont {Perrier},\ and\
  \citenamefont {Riotto}}]{Kehagias:2013rpa}%
  \BibitemOpen
  \bibfield  {author} {\bibinfo {author} {\bibfnamefont {A.}~\bibnamefont
  {Kehagias}}, \bibinfo {author} {\bibfnamefont {J.}~\bibnamefont {Nore\~na}},
  \bibinfo {author} {\bibfnamefont {H.}~\bibnamefont {Perrier}}, \ and\
  \bibinfo {author} {\bibfnamefont {A.}~\bibnamefont {Riotto}},\ }\href
  {\doibase 10.1016/j.nuclphysb.2014.03.020} {\bibfield  {journal} {\bibinfo
  {journal} {Nucl. Phys. B}\ }\textbf {\bibinfo {volume} {883}},\ \bibinfo
  {pages} {83} (\bibinfo {year} {2014})},\ \Eprint
  {http://arxiv.org/abs/1311.0786} {arXiv:1311.0786 [astro-ph.CO]} \BibitemShut
  {NoStop}%
\bibitem [{\citenamefont {Simonovi\'c}(2014)}]{Simonovic:2014yna}%
  \BibitemOpen
  \bibfield  {author} {\bibinfo {author} {\bibfnamefont {M.}~\bibnamefont
  {Simonovi\'c}},\ }\emph {\bibinfo {title} {{Cosmological Consistency
  Relations}}},\ \href@noop {} {Ph.D. thesis},\ \bibinfo  {school} {SISSA,
  Trieste} (\bibinfo {year} {2014})\BibitemShut {NoStop}%
\bibitem [{\citenamefont {Kehagias}\ \emph {et~al.}(2015)\citenamefont
  {Kehagias}, \citenamefont {Moradinezhad~Dizgah}, \citenamefont {Nore\~na},
  \citenamefont {Perrier},\ and\ \citenamefont {Riotto}}]{Kehagias:2015tda}%
  \BibitemOpen
  \bibfield  {author} {\bibinfo {author} {\bibfnamefont {A.}~\bibnamefont
  {Kehagias}}, \bibinfo {author} {\bibfnamefont {A.}~\bibnamefont
  {Moradinezhad~Dizgah}}, \bibinfo {author} {\bibfnamefont {J.}~\bibnamefont
  {Nore\~na}}, \bibinfo {author} {\bibfnamefont {H.}~\bibnamefont {Perrier}}, \
  and\ \bibinfo {author} {\bibfnamefont {A.}~\bibnamefont {Riotto}},\ }\href
  {\doibase 10.1088/1475-7516/2015/08/018} {\bibfield  {journal} {\bibinfo
  {journal} {JCAP}\ }\textbf {\bibinfo {volume} {08}},\ \bibinfo {pages} {018}
  (\bibinfo {year} {2015})},\ \Eprint {http://arxiv.org/abs/1503.04467}
  {arXiv:1503.04467 [astro-ph.CO]} \BibitemShut {NoStop}%
\bibitem [{\citenamefont {Creminelli}\ \emph
  {et~al.}(2014{\natexlab{b}})\citenamefont {Creminelli}, \citenamefont
  {Gleyzes}, \citenamefont {Hui}, \citenamefont {Simonovi\'c},\ and\
  \citenamefont {Vernizzi}}]{Creminelli:2013nua}%
  \BibitemOpen
  \bibfield  {author} {\bibinfo {author} {\bibfnamefont {P.}~\bibnamefont
  {Creminelli}}, \bibinfo {author} {\bibfnamefont {J.}~\bibnamefont {Gleyzes}},
  \bibinfo {author} {\bibfnamefont {L.}~\bibnamefont {Hui}}, \bibinfo {author}
  {\bibfnamefont {M.}~\bibnamefont {Simonovi\'c}}, \ and\ \bibinfo {author}
  {\bibfnamefont {F.}~\bibnamefont {Vernizzi}},\ }\href {\doibase
  10.1088/1475-7516/2014/06/009} {\bibfield  {journal} {\bibinfo  {journal}
  {JCAP}\ }\textbf {\bibinfo {volume} {06}},\ \bibinfo {pages} {009} (\bibinfo
  {year} {2014}{\natexlab{b}})},\ \Eprint {http://arxiv.org/abs/1312.6074}
  {arXiv:1312.6074 [astro-ph.CO]} \BibitemShut {NoStop}%
\bibitem [{\citenamefont {Lyth}\ and\ \citenamefont
  {Wands}(2002)}]{Lyth:2001nq}%
  \BibitemOpen
  \bibfield  {author} {\bibinfo {author} {\bibfnamefont {D.~H.}\ \bibnamefont
  {Lyth}}\ and\ \bibinfo {author} {\bibfnamefont {D.}~\bibnamefont {Wands}},\
  }\href {\doibase 10.1016/S0370-2693(01)01366-1} {\bibfield  {journal}
  {\bibinfo  {journal} {Phys. Lett. B}\ }\textbf {\bibinfo {volume} {524}},\
  \bibinfo {pages} {5} (\bibinfo {year} {2002})},\ \Eprint
  {http://arxiv.org/abs/hep-ph/0110002} {arXiv:hep-ph/0110002} \BibitemShut
  {NoStop}%
\bibitem [{\citenamefont {Zaldarriaga}(2004)}]{Zaldarriaga:2003my}%
  \BibitemOpen
  \bibfield  {author} {\bibinfo {author} {\bibfnamefont {M.}~\bibnamefont
  {Zaldarriaga}},\ }\href {\doibase 10.1103/PhysRevD.69.043508} {\bibfield
  {journal} {\bibinfo  {journal} {Phys. Rev. D}\ }\textbf {\bibinfo {volume}
  {69}},\ \bibinfo {pages} {043508} (\bibinfo {year} {2004})},\ \Eprint
  {http://arxiv.org/abs/astro-ph/0306006} {arXiv:astro-ph/0306006} \BibitemShut
  {NoStop}%
\bibitem [{\citenamefont {Chen}\ and\ \citenamefont
  {Wang}(2010)}]{Chen:2009zp}%
  \BibitemOpen
  \bibfield  {author} {\bibinfo {author} {\bibfnamefont {X.}~\bibnamefont
  {Chen}}\ and\ \bibinfo {author} {\bibfnamefont {Y.}~\bibnamefont {Wang}},\
  }\href {\doibase 10.1088/1475-7516/2010/04/027} {\bibfield  {journal}
  {\bibinfo  {journal} {JCAP}\ }\textbf {\bibinfo {volume} {04}},\ \bibinfo
  {pages} {027} (\bibinfo {year} {2010})},\ \Eprint
  {http://arxiv.org/abs/0911.3380} {arXiv:0911.3380 [hep-th]} \BibitemShut
  {NoStop}%
\bibitem [{\citenamefont {Baumann}\ and\ \citenamefont
  {Green}(2012)}]{Baumann:2011nk}%
  \BibitemOpen
  \bibfield  {author} {\bibinfo {author} {\bibfnamefont {D.}~\bibnamefont
  {Baumann}}\ and\ \bibinfo {author} {\bibfnamefont {D.}~\bibnamefont
  {Green}},\ }\href {\doibase 10.1103/PhysRevD.85.103520} {\bibfield  {journal}
  {\bibinfo  {journal} {Phys. Rev. D}\ }\textbf {\bibinfo {volume} {85}},\
  \bibinfo {pages} {103520} (\bibinfo {year} {2012})},\ \Eprint
  {http://arxiv.org/abs/1109.0292} {arXiv:1109.0292 [hep-th]} \BibitemShut
  {NoStop}%
\bibitem [{\citenamefont {Noumi}\ \emph {et~al.}(2013)\citenamefont {Noumi},
  \citenamefont {Yamaguchi},\ and\ \citenamefont {Yokoyama}}]{Noumi:2012vr}%
  \BibitemOpen
  \bibfield  {author} {\bibinfo {author} {\bibfnamefont {T.}~\bibnamefont
  {Noumi}}, \bibinfo {author} {\bibfnamefont {M.}~\bibnamefont {Yamaguchi}}, \
  and\ \bibinfo {author} {\bibfnamefont {D.}~\bibnamefont {Yokoyama}},\ }\href
  {\doibase 10.1007/JHEP06(2013)051} {\bibfield  {journal} {\bibinfo  {journal}
  {JHEP}\ }\textbf {\bibinfo {volume} {06}},\ \bibinfo {pages} {051} (\bibinfo
  {year} {2013})},\ \Eprint {http://arxiv.org/abs/1211.1624} {arXiv:1211.1624
  [hep-th]} \BibitemShut {NoStop}%
\bibitem [{\citenamefont {Desjacques}\ \emph {et~al.}(2018)\citenamefont
  {Desjacques}, \citenamefont {Jeong},\ and\ \citenamefont
  {Schmidt}}]{Desjacques:2016bnm}%
  \BibitemOpen
  \bibfield  {author} {\bibinfo {author} {\bibfnamefont {V.}~\bibnamefont
  {Desjacques}}, \bibinfo {author} {\bibfnamefont {D.}~\bibnamefont {Jeong}}, \
  and\ \bibinfo {author} {\bibfnamefont {F.}~\bibnamefont {Schmidt}},\ }\href
  {\doibase 10.1016/j.physrep.2017.12.002} {\bibfield  {journal} {\bibinfo
  {journal} {Phys. Rept.}\ }\textbf {\bibinfo {volume} {733}},\ \bibinfo
  {pages} {1} (\bibinfo {year} {2018})},\ \Eprint
  {http://arxiv.org/abs/1611.09787} {arXiv:1611.09787 [astro-ph.CO]}
  \BibitemShut {NoStop}%
\bibitem [{\citenamefont {Komatsu}\ and\ \citenamefont
  {Spergel}(2001)}]{Komatsu:2001rj}%
  \BibitemOpen
  \bibfield  {author} {\bibinfo {author} {\bibfnamefont {E.}~\bibnamefont
  {Komatsu}}\ and\ \bibinfo {author} {\bibfnamefont {D.~N.}\ \bibnamefont
  {Spergel}},\ }\href {\doibase 10.1103/PhysRevD.63.063002} {\bibfield
  {journal} {\bibinfo  {journal} {Phys. Rev. D}\ }\textbf {\bibinfo {volume}
  {63}},\ \bibinfo {pages} {063002} (\bibinfo {year} {2001})},\ \Eprint
  {http://arxiv.org/abs/astro-ph/0005036} {arXiv:astro-ph/0005036} \BibitemShut
  {NoStop}%
\bibitem [{\citenamefont {Scoccimarro}\ \emph {et~al.}(2012)\citenamefont
  {Scoccimarro}, \citenamefont {Hui}, \citenamefont {Manera},\ and\
  \citenamefont {Chan}}]{Scoccimarro:2011pz}%
  \BibitemOpen
  \bibfield  {author} {\bibinfo {author} {\bibfnamefont {R.}~\bibnamefont
  {Scoccimarro}}, \bibinfo {author} {\bibfnamefont {L.}~\bibnamefont {Hui}},
  \bibinfo {author} {\bibfnamefont {M.}~\bibnamefont {Manera}}, \ and\ \bibinfo
  {author} {\bibfnamefont {K.~C.}\ \bibnamefont {Chan}},\ }\href {\doibase
  10.1103/PhysRevD.85.083002} {\bibfield  {journal} {\bibinfo  {journal} {Phys.
  Rev. D}\ }\textbf {\bibinfo {volume} {85}},\ \bibinfo {pages} {083002}
  (\bibinfo {year} {2012})},\ \Eprint {http://arxiv.org/abs/1108.5512}
  {arXiv:1108.5512 [astro-ph.CO]} \BibitemShut {NoStop}%
\bibitem [{\citenamefont {Slosar}\ \emph {et~al.}(2008)\citenamefont {Slosar},
  \citenamefont {Hirata}, \citenamefont {Seljak}, \citenamefont {Ho},\ and\
  \citenamefont {Padmanabhan}}]{Slosar:2008hx}%
  \BibitemOpen
  \bibfield  {author} {\bibinfo {author} {\bibfnamefont {A.}~\bibnamefont
  {Slosar}}, \bibinfo {author} {\bibfnamefont {C.}~\bibnamefont {Hirata}},
  \bibinfo {author} {\bibfnamefont {U.}~\bibnamefont {Seljak}}, \bibinfo
  {author} {\bibfnamefont {S.}~\bibnamefont {Ho}}, \ and\ \bibinfo {author}
  {\bibfnamefont {N.}~\bibnamefont {Padmanabhan}},\ }\href {\doibase
  10.1088/1475-7516/2008/08/031} {\bibfield  {journal} {\bibinfo  {journal}
  {JCAP}\ }\textbf {\bibinfo {volume} {08}},\ \bibinfo {pages} {031} (\bibinfo
  {year} {2008})},\ \Eprint {http://arxiv.org/abs/0805.3580} {arXiv:0805.3580
  [astro-ph]} \BibitemShut {NoStop}%
\bibitem [{\citenamefont {Chiang}(2017)}]{Chiang:2017jnm}%
  \BibitemOpen
  \bibfield  {author} {\bibinfo {author} {\bibfnamefont {C.-T.}\ \bibnamefont
  {Chiang}},\ }\href {\doibase 10.1103/PhysRevD.95.123517} {\bibfield
  {journal} {\bibinfo  {journal} {Phys. Rev. D}\ }\textbf {\bibinfo {volume}
  {95}},\ \bibinfo {pages} {123517} (\bibinfo {year} {2017})},\ \Eprint
  {http://arxiv.org/abs/1701.03374} {arXiv:1701.03374 [astro-ph.CO]}
  \BibitemShut {NoStop}%
\bibitem [{\citenamefont {Giri}\ \emph {et~al.}(2022)\citenamefont {Giri},
  \citenamefont {M\"unchmeyer},\ and\ \citenamefont {Smith}}]{Giri:2022nzt}%
  \BibitemOpen
  \bibfield  {author} {\bibinfo {author} {\bibfnamefont {U.}~\bibnamefont
  {Giri}}, \bibinfo {author} {\bibfnamefont {M.}~\bibnamefont {M\"unchmeyer}},
  \ and\ \bibinfo {author} {\bibfnamefont {K.~M.}\ \bibnamefont {Smith}},\
  }\href@noop {} {\  (\bibinfo {year} {2022})},\ \Eprint
  {http://arxiv.org/abs/2205.12964} {arXiv:2205.12964 [astro-ph.CO]}
  \BibitemShut {NoStop}%
\bibitem [{\citenamefont {Baldauf}\ \emph {et~al.}(2011)\citenamefont
  {Baldauf}, \citenamefont {Seljak},\ and\ \citenamefont
  {Senatore}}]{Baldauf:2010vn}%
  \BibitemOpen
  \bibfield  {author} {\bibinfo {author} {\bibfnamefont {T.}~\bibnamefont
  {Baldauf}}, \bibinfo {author} {\bibfnamefont {U.}~\bibnamefont {Seljak}}, \
  and\ \bibinfo {author} {\bibfnamefont {L.}~\bibnamefont {Senatore}},\ }\href
  {\doibase 10.1088/1475-7516/2011/04/006} {\bibfield  {journal} {\bibinfo
  {journal} {JCAP}\ }\textbf {\bibinfo {volume} {04}},\ \bibinfo {pages} {006}
  (\bibinfo {year} {2011})},\ \Eprint {http://arxiv.org/abs/1011.1513}
  {arXiv:1011.1513 [astro-ph.CO]} \BibitemShut {NoStop}%
\bibitem [{\citenamefont {Horn}\ \emph {et~al.}(2014)\citenamefont {Horn},
  \citenamefont {Hui},\ and\ \citenamefont {Xiao}}]{Horn:2014rta}%
  \BibitemOpen
  \bibfield  {author} {\bibinfo {author} {\bibfnamefont {B.}~\bibnamefont
  {Horn}}, \bibinfo {author} {\bibfnamefont {L.}~\bibnamefont {Hui}}, \ and\
  \bibinfo {author} {\bibfnamefont {X.}~\bibnamefont {Xiao}},\ }\href {\doibase
  10.1088/1475-7516/2014/09/044} {\bibfield  {journal} {\bibinfo  {journal}
  {JCAP}\ }\textbf {\bibinfo {volume} {09}},\ \bibinfo {pages} {044} (\bibinfo
  {year} {2014})},\ \Eprint {http://arxiv.org/abs/1406.0842} {arXiv:1406.0842
  [hep-th]} \BibitemShut {NoStop}%
\bibitem [{\citenamefont {Nishimichi}\ and\ \citenamefont
  {Valageas}(2014)}]{Nishimichi:2014jna}%
  \BibitemOpen
  \bibfield  {author} {\bibinfo {author} {\bibfnamefont {T.}~\bibnamefont
  {Nishimichi}}\ and\ \bibinfo {author} {\bibfnamefont {P.}~\bibnamefont
  {Valageas}},\ }\href {\doibase 10.1103/PhysRevD.90.023546} {\bibfield
  {journal} {\bibinfo  {journal} {Phys. Rev. D}\ }\textbf {\bibinfo {volume}
  {90}},\ \bibinfo {pages} {023546} (\bibinfo {year} {2014})},\ \Eprint
  {http://arxiv.org/abs/1402.3293} {arXiv:1402.3293 [astro-ph.CO]} \BibitemShut
  {NoStop}%
\bibitem [{\citenamefont {Lewis}(2011)}]{Lewis:2011au}%
  \BibitemOpen
  \bibfield  {author} {\bibinfo {author} {\bibfnamefont {A.}~\bibnamefont
  {Lewis}},\ }\href {\doibase 10.1088/1475-7516/2011/10/026} {\bibfield
  {journal} {\bibinfo  {journal} {JCAP}\ }\textbf {\bibinfo {volume} {10}},\
  \bibinfo {pages} {026} (\bibinfo {year} {2011})},\ \Eprint
  {http://arxiv.org/abs/1107.5431} {arXiv:1107.5431 [astro-ph.CO]} \BibitemShut
  {NoStop}%
\bibitem [{\citenamefont {Takahashi}\ \emph {et~al.}(2012)\citenamefont
  {Takahashi}, \citenamefont {Sato}, \citenamefont {Nishimichi}, \citenamefont
  {Taruya},\ and\ \citenamefont {Oguri}}]{Takahashi:2012em}%
  \BibitemOpen
  \bibfield  {author} {\bibinfo {author} {\bibfnamefont {R.}~\bibnamefont
  {Takahashi}}, \bibinfo {author} {\bibfnamefont {M.}~\bibnamefont {Sato}},
  \bibinfo {author} {\bibfnamefont {T.}~\bibnamefont {Nishimichi}}, \bibinfo
  {author} {\bibfnamefont {A.}~\bibnamefont {Taruya}}, \ and\ \bibinfo {author}
  {\bibfnamefont {M.}~\bibnamefont {Oguri}},\ }\href {\doibase
  10.1088/0004-637X/761/2/152} {\bibfield  {journal} {\bibinfo  {journal}
  {Astrophys. J.}\ }\textbf {\bibinfo {volume} {761}},\ \bibinfo {pages} {152}
  (\bibinfo {year} {2012})},\ \Eprint {http://arxiv.org/abs/1208.2701}
  {arXiv:1208.2701 [astro-ph.CO]} \BibitemShut {NoStop}%
\bibitem [{\citenamefont {Villaescusa-Navarro}\ \emph
  {et~al.}(2020)\citenamefont {Villaescusa-Navarro} \emph
  {et~al.}}]{Villaescusa-Navarro:2019bje}%
  \BibitemOpen
  \bibfield  {author} {\bibinfo {author} {\bibfnamefont {F.}~\bibnamefont
  {Villaescusa-Navarro}} \emph {et~al.},\ }\href {\doibase
  10.3847/1538-4365/ab9d82} {\bibfield  {journal} {\bibinfo  {journal}
  {Astrophys. J. Suppl.}\ }\textbf {\bibinfo {volume} {250}},\ \bibinfo {pages}
  {2} (\bibinfo {year} {2020})},\ \Eprint {http://arxiv.org/abs/1909.05273}
  {arXiv:1909.05273 [astro-ph.CO]} \BibitemShut {NoStop}%
\bibitem [{\citenamefont {Sellentin}\ and\ \citenamefont
  {Heavens}(2016)}]{Sellentin:2015waz}%
  \BibitemOpen
  \bibfield  {author} {\bibinfo {author} {\bibfnamefont {E.}~\bibnamefont
  {Sellentin}}\ and\ \bibinfo {author} {\bibfnamefont {A.~F.}\ \bibnamefont
  {Heavens}},\ }\href {\doibase 10.1093/mnrasl/slv190} {\bibfield  {journal}
  {\bibinfo  {journal} {Mon. Not. Roy. Astron. Soc.}\ }\textbf {\bibinfo
  {volume} {456}},\ \bibinfo {pages} {L132} (\bibinfo {year} {2016})},\ \Eprint
  {http://arxiv.org/abs/1511.05969} {arXiv:1511.05969 [astro-ph.CO]}
  \BibitemShut {NoStop}%
\bibitem [{\citenamefont {Feroz}\ \emph {et~al.}(2009)\citenamefont {Feroz},
  \citenamefont {Hobson},\ and\ \citenamefont {Bridges}}]{Feroz:2008xx}%
  \BibitemOpen
  \bibfield  {author} {\bibinfo {author} {\bibfnamefont {F.}~\bibnamefont
  {Feroz}}, \bibinfo {author} {\bibfnamefont {M.~P.}\ \bibnamefont {Hobson}}, \
  and\ \bibinfo {author} {\bibfnamefont {M.}~\bibnamefont {Bridges}},\ }\href
  {\doibase 10.1111/j.1365-2966.2009.14548.x} {\bibfield  {journal} {\bibinfo
  {journal} {Mon. Not. Roy. Astron. Soc.}\ }\textbf {\bibinfo {volume} {398}},\
  \bibinfo {pages} {1601} (\bibinfo {year} {2009})},\ \Eprint
  {http://arxiv.org/abs/0809.3437} {arXiv:0809.3437 [astro-ph]} \BibitemShut
  {NoStop}%
\bibitem [{\citenamefont {Coulton}\ \emph {et~al.}(2022)\citenamefont
  {Coulton}, \citenamefont {Villaescusa-Navarro}, \citenamefont {Jamieson},
  \citenamefont {Baldi}, \citenamefont {Jung}, \citenamefont {Karagiannis},
  \citenamefont {Liguori}, \citenamefont {Verde},\ and\ \citenamefont
  {Wandelt}}]{Coulton:2022rir}%
  \BibitemOpen
  \bibfield  {author} {\bibinfo {author} {\bibfnamefont {W.~R.}\ \bibnamefont
  {Coulton}}, \bibinfo {author} {\bibfnamefont {F.}~\bibnamefont
  {Villaescusa-Navarro}}, \bibinfo {author} {\bibfnamefont {D.}~\bibnamefont
  {Jamieson}}, \bibinfo {author} {\bibfnamefont {M.}~\bibnamefont {Baldi}},
  \bibinfo {author} {\bibfnamefont {G.}~\bibnamefont {Jung}}, \bibinfo {author}
  {\bibfnamefont {D.}~\bibnamefont {Karagiannis}}, \bibinfo {author}
  {\bibfnamefont {M.}~\bibnamefont {Liguori}}, \bibinfo {author} {\bibfnamefont
  {L.}~\bibnamefont {Verde}}, \ and\ \bibinfo {author} {\bibfnamefont {B.~D.}\
  \bibnamefont {Wandelt}},\ }\href@noop {} {\  (\bibinfo {year} {2022})},\
  \Eprint {http://arxiv.org/abs/2206.15450} {arXiv:2206.15450 [astro-ph.CO]}
  \BibitemShut {NoStop}%
\bibitem [{\citenamefont {Fl\"oss}\ \emph {et~al.}(2022)\citenamefont
  {Fl\"oss}, \citenamefont {Biagetti},\ and\ \citenamefont
  {Meerburg}}]{Floss:2022wkq}%
  \BibitemOpen
  \bibfield  {author} {\bibinfo {author} {\bibfnamefont {T.}~\bibnamefont
  {Fl\"oss}}, \bibinfo {author} {\bibfnamefont {M.}~\bibnamefont {Biagetti}}, \
  and\ \bibinfo {author} {\bibfnamefont {P.~D.}\ \bibnamefont {Meerburg}},\
  }\href@noop {} {\  (\bibinfo {year} {2022})},\ \Eprint
  {http://arxiv.org/abs/2206.10458} {arXiv:2206.10458 [astro-ph.CO]}
  \BibitemShut {NoStop}%
\bibitem [{\citenamefont {Biagetti}\ \emph {et~al.}(2022)\citenamefont
  {Biagetti}, \citenamefont {Castiblanco}, \citenamefont {Nore\~na},\ and\
  \citenamefont {Sefusatti}}]{Biagetti:2021tua}%
  \BibitemOpen
  \bibfield  {author} {\bibinfo {author} {\bibfnamefont {M.}~\bibnamefont
  {Biagetti}}, \bibinfo {author} {\bibfnamefont {L.}~\bibnamefont
  {Castiblanco}}, \bibinfo {author} {\bibfnamefont {J.}~\bibnamefont
  {Nore\~na}}, \ and\ \bibinfo {author} {\bibfnamefont {E.}~\bibnamefont
  {Sefusatti}},\ }\href {\doibase 10.1088/1475-7516/2022/09/009} {\bibfield
  {journal} {\bibinfo  {journal} {JCAP}\ }\textbf {\bibinfo {volume} {09}},\
  \bibinfo {pages} {009} (\bibinfo {year} {2022})},\ \Eprint
  {http://arxiv.org/abs/2111.05887} {arXiv:2111.05887 [astro-ph.CO]}
  \BibitemShut {NoStop}%
\bibitem [{\citenamefont {Aghamousa}\ \emph {et~al.}(2016)\citenamefont
  {Aghamousa} \emph {et~al.}}]{DESI:2016fyo}%
  \BibitemOpen
  \bibfield  {author} {\bibinfo {author} {\bibfnamefont {A.}~\bibnamefont
  {Aghamousa}} \emph {et~al.} (\bibinfo {collaboration} {DESI}),\ }\href@noop
  {} {\  (\bibinfo {year} {2016})},\ \Eprint {http://arxiv.org/abs/1611.00036}
  {arXiv:1611.00036 [astro-ph.IM]} \BibitemShut {NoStop}%
\bibitem [{\citenamefont {Levi}\ \emph {et~al.}(2013)\citenamefont {Levi} \emph
  {et~al.}}]{DESI:2013agm}%
  \BibitemOpen
  \bibfield  {author} {\bibinfo {author} {\bibfnamefont {M.}~\bibnamefont
  {Levi}} \emph {et~al.} (\bibinfo {collaboration} {DESI}),\ }\href@noop {} {\
  (\bibinfo {year} {2013})},\ \Eprint {http://arxiv.org/abs/1308.0847}
  {arXiv:1308.0847 [astro-ph.CO]} \BibitemShut {NoStop}%
\bibitem [{\citenamefont {Abbott}\ \emph {et~al.}(2005)\citenamefont {Abbott}
  \emph {et~al.}}]{DES:2005dhi}%
  \BibitemOpen
  \bibfield  {author} {\bibinfo {author} {\bibfnamefont {T.}~\bibnamefont
  {Abbott}} \emph {et~al.} (\bibinfo {collaboration} {DES}),\ }\href@noop {} {\
   (\bibinfo {year} {2005})},\ \Eprint {http://arxiv.org/abs/astro-ph/0510346}
  {arXiv:astro-ph/0510346} \BibitemShut {NoStop}%
\bibitem [{\citenamefont {Abell}\ \emph {et~al.}(2009)\citenamefont {Abell}
  \emph {et~al.}}]{LSSTScience:2009jmu}%
  \BibitemOpen
  \bibfield  {author} {\bibinfo {author} {\bibfnamefont {P.~A.}\ \bibnamefont
  {Abell}} \emph {et~al.} (\bibinfo {collaboration} {LSST Science, LSST
  Project}),\ }\href@noop {} {\  (\bibinfo {year} {2009})},\ \Eprint
  {http://arxiv.org/abs/0912.0201} {arXiv:0912.0201 [astro-ph.IM]} \BibitemShut
  {NoStop}%
\bibitem [{\citenamefont {Kosowsky}(2003)}]{Kosowsky:2003smi}%
  \BibitemOpen
  \bibfield  {author} {\bibinfo {author} {\bibfnamefont {A.}~\bibnamefont
  {Kosowsky}},\ }\href {\doibase 10.1016/j.newar.2003.09.003} {\bibfield
  {journal} {\bibinfo  {journal} {New Astron. Rev.}\ }\textbf {\bibinfo
  {volume} {47}},\ \bibinfo {pages} {939} (\bibinfo {year} {2003})},\ \Eprint
  {http://arxiv.org/abs/astro-ph/0402234} {arXiv:astro-ph/0402234} \BibitemShut
  {NoStop}%
\bibitem [{\citenamefont {Ade}\ \emph {et~al.}(2019)\citenamefont {Ade} \emph
  {et~al.}}]{SimonsObservatory:2018koc}%
  \BibitemOpen
  \bibfield  {author} {\bibinfo {author} {\bibfnamefont {P.}~\bibnamefont
  {Ade}} \emph {et~al.} (\bibinfo {collaboration} {Simons Observatory}),\
  }\href {\doibase 10.1088/1475-7516/2019/02/056} {\bibfield  {journal}
  {\bibinfo  {journal} {JCAP}\ }\textbf {\bibinfo {volume} {02}},\ \bibinfo
  {pages} {056} (\bibinfo {year} {2019})},\ \Eprint
  {http://arxiv.org/abs/1808.07445} {arXiv:1808.07445 [astro-ph.CO]}
  \BibitemShut {NoStop}%
\bibitem [{\citenamefont {Abazajian}\ \emph {et~al.}(2016)\citenamefont
  {Abazajian} \emph {et~al.}}]{CMB-S4:2016ple}%
  \BibitemOpen
  \bibfield  {author} {\bibinfo {author} {\bibfnamefont {K.~N.}\ \bibnamefont
  {Abazajian}} \emph {et~al.} (\bibinfo {collaboration} {CMB-S4}),\ }\href@noop
  {} {\  (\bibinfo {year} {2016})},\ \Eprint {http://arxiv.org/abs/1610.02743}
  {arXiv:1610.02743 [astro-ph.CO]} \BibitemShut {NoStop}%
\bibitem [{\citenamefont {Barreira}(2022)}]{Barreira:2022sey}%
  \BibitemOpen
  \bibfield  {author} {\bibinfo {author} {\bibfnamefont {A.}~\bibnamefont
  {Barreira}},\ }\href@noop {} {\  (\bibinfo {year} {2022})},\ \Eprint
  {http://arxiv.org/abs/2205.05673} {arXiv:2205.05673 [astro-ph.CO]}
  \BibitemShut {NoStop}%
\bibitem [{\citenamefont {Reid}\ \emph {et~al.}(2016)\citenamefont {Reid} \emph
  {et~al.}}]{Reid:2015gra}%
  \BibitemOpen
  \bibfield  {author} {\bibinfo {author} {\bibfnamefont {B.}~\bibnamefont
  {Reid}} \emph {et~al.},\ }\href {\doibase 10.1093/mnras/stv2382} {\bibfield
  {journal} {\bibinfo  {journal} {Mon. Not. Roy. Astron. Soc.}\ }\textbf
  {\bibinfo {volume} {455}},\ \bibinfo {pages} {1553} (\bibinfo {year}
  {2016})},\ \Eprint {http://arxiv.org/abs/1509.06529} {arXiv:1509.06529
  [astro-ph.CO]} \BibitemShut {NoStop}%
\bibitem [{\citenamefont {Amendola}\ \emph {et~al.}(2018)\citenamefont
  {Amendola} \emph {et~al.}}]{Amendola:2016saw}%
  \BibitemOpen
  \bibfield  {author} {\bibinfo {author} {\bibfnamefont {L.}~\bibnamefont
  {Amendola}} \emph {et~al.},\ }\href {\doibase 10.1007/s41114-017-0010-3}
  {\bibfield  {journal} {\bibinfo  {journal} {Living Rev. Rel.}\ }\textbf
  {\bibinfo {volume} {21}},\ \bibinfo {pages} {2} (\bibinfo {year} {2018})},\
  \Eprint {http://arxiv.org/abs/1606.00180} {arXiv:1606.00180 [astro-ph.CO]}
  \BibitemShut {NoStop}%
\bibitem [{\citenamefont {Dor\'e}\ \emph {et~al.}(2014)\citenamefont {Dor\'e}
  \emph {et~al.}}]{Dore:2014cca}%
  \BibitemOpen
  \bibfield  {author} {\bibinfo {author} {\bibfnamefont {O.}~\bibnamefont
  {Dor\'e}} \emph {et~al.},\ }\href@noop {} {\  (\bibinfo {year} {2014})},\
  \Eprint {http://arxiv.org/abs/1412.4872} {arXiv:1412.4872 [astro-ph.CO]}
  \BibitemShut {NoStop}%
\bibitem [{\citenamefont {Cabass}\ \emph {et~al.}(2022)\citenamefont {Cabass},
  \citenamefont {Ivanov}, \citenamefont {Philcox}, \citenamefont
  {Simonovi\'c},\ and\ \citenamefont {Zaldarriaga}}]{Cabass:2022ymb}%
  \BibitemOpen
  \bibfield  {author} {\bibinfo {author} {\bibfnamefont {G.}~\bibnamefont
  {Cabass}}, \bibinfo {author} {\bibfnamefont {M.~M.}\ \bibnamefont {Ivanov}},
  \bibinfo {author} {\bibfnamefont {O.~H.~E.}\ \bibnamefont {Philcox}},
  \bibinfo {author} {\bibfnamefont {M.}~\bibnamefont {Simonovi\'c}}, \ and\
  \bibinfo {author} {\bibfnamefont {M.}~\bibnamefont {Zaldarriaga}},\ }\href
  {\doibase 10.1103/PhysRevD.106.043506} {\bibfield  {journal} {\bibinfo
  {journal} {Phys. Rev. D}\ }\textbf {\bibinfo {volume} {106}},\ \bibinfo
  {pages} {043506} (\bibinfo {year} {2022})},\ \Eprint
  {http://arxiv.org/abs/2204.01781} {arXiv:2204.01781 [astro-ph.CO]}
  \BibitemShut {NoStop}%
\bibitem [{\citenamefont {D'Amico}\ \emph {et~al.}(2022)\citenamefont
  {D'Amico}, \citenamefont {Lewandowski}, \citenamefont {Senatore},\ and\
  \citenamefont {Zhang}}]{DAmico:2022gki}%
  \BibitemOpen
  \bibfield  {author} {\bibinfo {author} {\bibfnamefont {G.}~\bibnamefont
  {D'Amico}}, \bibinfo {author} {\bibfnamefont {M.}~\bibnamefont
  {Lewandowski}}, \bibinfo {author} {\bibfnamefont {L.}~\bibnamefont
  {Senatore}}, \ and\ \bibinfo {author} {\bibfnamefont {P.}~\bibnamefont
  {Zhang}},\ }\href@noop {} {\  (\bibinfo {year} {2022})},\ \Eprint
  {http://arxiv.org/abs/2201.11518} {arXiv:2201.11518 [astro-ph.CO]}
  \BibitemShut {NoStop}%
\bibitem [{\citenamefont {Andrews}\ \emph {et~al.}(2022)\citenamefont
  {Andrews}, \citenamefont {Jasche}, \citenamefont {Lavaux},\ and\
  \citenamefont {Schmidt}}]{Andrews:2022nvv}%
  \BibitemOpen
  \bibfield  {author} {\bibinfo {author} {\bibfnamefont {A.}~\bibnamefont
  {Andrews}}, \bibinfo {author} {\bibfnamefont {J.}~\bibnamefont {Jasche}},
  \bibinfo {author} {\bibfnamefont {G.}~\bibnamefont {Lavaux}}, \ and\ \bibinfo
  {author} {\bibfnamefont {F.}~\bibnamefont {Schmidt}},\ }\href@noop {} {\
  (\bibinfo {year} {2022})},\ \Eprint {http://arxiv.org/abs/2203.08838}
  {arXiv:2203.08838 [astro-ph.CO]} \BibitemShut {NoStop}%
\bibitem [{\citenamefont {Akrami}\ \emph {et~al.}(2020)\citenamefont {Akrami}
  \emph {et~al.}}]{Planck:2019kim}%
  \BibitemOpen
  \bibfield  {author} {\bibinfo {author} {\bibfnamefont {Y.}~\bibnamefont
  {Akrami}} \emph {et~al.} (\bibinfo {collaboration} {Planck}),\ }\href
  {\doibase 10.1051/0004-6361/201935891} {\bibfield  {journal} {\bibinfo
  {journal} {Astron. Astrophys.}\ }\textbf {\bibinfo {volume} {641}},\ \bibinfo
  {pages} {A9} (\bibinfo {year} {2020})},\ \Eprint
  {http://arxiv.org/abs/1905.05697} {arXiv:1905.05697 [astro-ph.CO]}
  \BibitemShut {NoStop}%
\bibitem [{\citenamefont {Bernardeau}\ \emph {et~al.}(2002)\citenamefont
  {Bernardeau}, \citenamefont {Colombi}, \citenamefont {Gaztanaga},\ and\
  \citenamefont {Scoccimarro}}]{Bernardeau:2001qr}%
  \BibitemOpen
  \bibfield  {author} {\bibinfo {author} {\bibfnamefont {F.}~\bibnamefont
  {Bernardeau}}, \bibinfo {author} {\bibfnamefont {S.}~\bibnamefont {Colombi}},
  \bibinfo {author} {\bibfnamefont {E.}~\bibnamefont {Gaztanaga}}, \ and\
  \bibinfo {author} {\bibfnamefont {R.}~\bibnamefont {Scoccimarro}},\ }\href
  {\doibase 10.1016/S0370-1573(02)00135-7} {\bibfield  {journal} {\bibinfo
  {journal} {Phys. Rept.}\ }\textbf {\bibinfo {volume} {367}},\ \bibinfo
  {pages} {1} (\bibinfo {year} {2002})},\ \Eprint
  {http://arxiv.org/abs/astro-ph/0112551} {arXiv:astro-ph/0112551} \BibitemShut
  {NoStop}%
\bibitem [{\citenamefont {Baumann}\ \emph {et~al.}(2012)\citenamefont
  {Baumann}, \citenamefont {Nicolis}, \citenamefont {Senatore},\ and\
  \citenamefont {Zaldarriaga}}]{Baumann:2010tm}%
  \BibitemOpen
  \bibfield  {author} {\bibinfo {author} {\bibfnamefont {D.}~\bibnamefont
  {Baumann}}, \bibinfo {author} {\bibfnamefont {A.}~\bibnamefont {Nicolis}},
  \bibinfo {author} {\bibfnamefont {L.}~\bibnamefont {Senatore}}, \ and\
  \bibinfo {author} {\bibfnamefont {M.}~\bibnamefont {Zaldarriaga}},\ }\href
  {\doibase 10.1088/1475-7516/2012/07/051} {\bibfield  {journal} {\bibinfo
  {journal} {JCAP}\ }\textbf {\bibinfo {volume} {07}},\ \bibinfo {pages} {051}
  (\bibinfo {year} {2012})},\ \Eprint {http://arxiv.org/abs/1004.2488}
  {arXiv:1004.2488 [astro-ph.CO]} \BibitemShut {NoStop}%
\bibitem [{\citenamefont {Carrasco}\ \emph {et~al.}(2012)\citenamefont
  {Carrasco}, \citenamefont {Hertzberg},\ and\ \citenamefont
  {Senatore}}]{Carrasco:2012cv}%
  \BibitemOpen
  \bibfield  {author} {\bibinfo {author} {\bibfnamefont {J.~J.~M.}\
  \bibnamefont {Carrasco}}, \bibinfo {author} {\bibfnamefont {M.~P.}\
  \bibnamefont {Hertzberg}}, \ and\ \bibinfo {author} {\bibfnamefont
  {L.}~\bibnamefont {Senatore}},\ }\href {\doibase 10.1007/JHEP09(2012)082}
  {\bibfield  {journal} {\bibinfo  {journal} {JHEP}\ }\textbf {\bibinfo
  {volume} {09}},\ \bibinfo {pages} {082} (\bibinfo {year} {2012})},\ \Eprint
  {http://arxiv.org/abs/1206.2926} {arXiv:1206.2926 [astro-ph.CO]} \BibitemShut
  {NoStop}%
\bibitem [{\citenamefont {Taruya}\ \emph {et~al.}(2008)\citenamefont {Taruya},
  \citenamefont {Koyama},\ and\ \citenamefont {Matsubara}}]{Taruya:2008pg}%
  \BibitemOpen
  \bibfield  {author} {\bibinfo {author} {\bibfnamefont {A.}~\bibnamefont
  {Taruya}}, \bibinfo {author} {\bibfnamefont {K.}~\bibnamefont {Koyama}}, \
  and\ \bibinfo {author} {\bibfnamefont {T.}~\bibnamefont {Matsubara}},\ }\href
  {\doibase 10.1103/PhysRevD.78.123534} {\bibfield  {journal} {\bibinfo
  {journal} {Phys. Rev. D}\ }\textbf {\bibinfo {volume} {78}},\ \bibinfo
  {pages} {123534} (\bibinfo {year} {2008})},\ \Eprint
  {http://arxiv.org/abs/0808.4085} {arXiv:0808.4085 [astro-ph]} \BibitemShut
  {NoStop}%
\end{thebibliography}%

\end{document}